\DocumentMetadata{
  lang=en-US,
  pdfversion=1.7,
  testphase={phase-III,table,title,math,firstaid}
}
\makeatletter
\@ifundefined{tagtool}{%
  \PackageError{acmart-arxiv-tagging}{Load DocumentMetadata with phase-III first}%
    {The source-tagging compatibility layer must precede documentclass.}%
}{%
  \let\ACMtag@startsection\@startsection
  \let\ACMtag@sect\@sect
  \let\ACMtag@ssect\@ssect
  \let\ACMtag@xsect\@xsect
  \AddToHook{class/acmart/after}[acmart-arxiv-tagging]{%
    \let\@startsection\ACMtag@startsection
    \def\@sect#1#2#3#4#5#6[#7]#8{%
      \setcounter{section@level}{#2}%
      \NR@gettitle{#7}%
      \ACMtag@sect{#1}{#2}{#3}{#4}{#5}{#6}[{#7}]{\Sectionformat{#8}{#2}}%
    }%
    \def\@ssect#1#2#3#4#5{%
      \NR@gettitle{#5}%
      \ACMtag@ssect{#1}{#2}{#3}{#4}{\Sectionformat{#5}{#1}}%
    }%
    \let\@xsect\ACMtag@xsect
    \patchcmd{\@mkauthors@i}{\noindent\unvbox}{\unvbox}
      {}{\PackageWarning{acmart-arxiv-tagging}{Author box patch did not match}}%
    \patchcmd{\@mkauthors@i}{\noindent\addresses}{\addresses}
      {}{\PackageWarning{acmart-arxiv-tagging}{Author content patch did not match}}%
    \patchcmd{\@mkteasers}{\noindent\unvbox}{\unvbox}
      {}{\PackageWarning{acmart-arxiv-tagging}{Teaser box patch did not match}}%
    \patchcmd{\@mkteasers}{\noindent\@Description@presentfalse}{\@Description@presentfalse}
      {}{\PackageWarning{acmart-arxiv-tagging}{Teaser content patch did not match}}%
    \renewcommand\textrightarrow{%
      \begingroup
      \ACMtag@artifactBegin\SuspendTagging{acmart-ccs-arrow}\MathCollectFalse
      $\rightarrow$%
      \ResumeTagging{acmart-ccs-arrow}\ACMtag@artifactEnd
      \endgroup}%
    \patchcmd{\@mktitle@i}
      {\@titlefont\noindent}
      {\@titlefont\tagtool{paratag=Title}\noindent}
      {}{\PackageWarning{acmart-arxiv-tagging}{Title patch did not match}}%
    \patchcmd{\@mkteasers}
      {\captionsetup{type=figure}}
      {\captionsetup{type=figure}\UseTaggingSocket{float/begin}}
      {}{\PackageWarning{acmart-arxiv-tagging}{Teaser begin patch did not match}}%
    \patchcmd{\@mkteasers}
      {\egroup\par}
      {\par\UseTaggingSocket{float/end}\egroup\par}
      {}{\PackageWarning{acmart-arxiv-tagging}{Teaser end patch did not match}}%
    \let\ACMtag@makecaption\caption@makecaption
    \let\ACMtag@captionSLC\caption@slc
    \long\def\caption@slc#1#2#3#4{%
      \SuspendTagging{acmart-caption-measure}%
      \ACMtag@captionSLC{#1}{#2}%
        {\ResumeTagging{acmart-caption-measure}#3}%
        {\ResumeTagging{acmart-caption-measure}#4}%
    }%
    \AddToHook{begindocument/end}[acmart-arxiv-tagging/caption]{%
      \long\def\@makecaption#1#2{%
        \UseTaggingSocket{caption/begin}{\@current@float@struct}%
        \ACMtag@makecaption{#1}{#2}%
        \UseTaggingSocket{caption/end}%
      }%
    }%
  }%
  \AddToHook{package/listings/after}[acmart-arxiv-tagging/listings]{%
    \ExplSyntaxOn
    \DeclareInstance{blockenv}{lstlisting}{display}
      {
        env-name = lstlisting,
        tag-name = verbatim,
        tag-class = ,
        tagging-recipe = standard,
        inner-level-counter = ,
        level-increase = false,
        setup-code = ,
        block-instance = displayblock,
        inner-instance = ,
        final-code = \tagtool{paratag=codeline},
        para-flattened = true
      }
    \ExplSyntaxOff
    \lstnewenvironment{taglstlisting}[2][]{%
      \MathCollectFalse
      \UseInstance{blockenv}{lstlisting}{}%
      \lst@TestEOLChar{#2}%
      \lstset{#1}%
      \csname\@lst @SetFirstNumber\endcsname
    }{%
      \@nobreakfalse
      \csname\@lst @SaveFirstNumber\endcsname
      \endblockenv
    }%
    \let\ACMtag@frameMakeBoxV\lst@frameMakeBoxV
    \def\lst@frameMakeBoxV#1#2#3{%
      \ACMtag@frameMakeBoxV{#1}{#2}{#3}%
      \setbox#1\hbox{%
        \ACMtag@artifactBegin\box#1\ACMtag@artifactEnd}%
    }%
    \let\ACMtag@frameH\lst@frameH
    \def\lst@frameH#1#2{%
      \ACMtag@frameH{#1}{#2}%
      \ifvoid\z@\else
        \setbox\z@\hbox{%
          \ACMtag@artifactBegin\box\z@\ACMtag@artifactEnd}%
      \fi
    }%
  }%
}
\ExplSyntaxOn
\cs_new_protected:Npn \ACMtag@artifactBegin { \pdf_bmc:n {Artifact} }
\cs_new_protected:Npn \ACMtag@artifactEnd { \pdf_emc: }
\AddToHook{begindocument/end}[acmart-arxiv-tagging/line-links]
  {
    \cs_new_eq:NN \ACMtag@savedLinkUrl \hyper@linkurl
    \cs_set_protected:Npn \hyper@linkurl #1 #2
      {
        \str_case_e:nnF {#2}
          {
            {https://www.reddit.com/r/Blind/comments/ohx5jq/those_who_live_alone_how_do_you_assemble_furniture/}
              {
                \ACMtag@savedLinkUrl
                  {\mbox{\Hurl{https://www.reddit.com/r/Blind/comments/ohx5jq/those_who_live_alone_}}}{#2}
                \allowbreak
                \ACMtag@savedLinkUrl
                  {\mbox{\Hurl{how_do_you_assemble_furniture/}}}{#2}
              }
            {https://doi.org/10.1145/3242587.3242604}
              {
                \ACMtag@savedLinkUrl
                  {\mbox{doi:\Hurl{10.1145/3242587.}}}{#2}
                \allowbreak
                \ACMtag@savedLinkUrl
                  {\mbox{\Hurl{3242604}}}{#2}
              }
          }
          {\ACMtag@savedLinkUrl{#1}{#2}}
      }
  }
\cs_new_protected:Npn \ACMTableRowSpan #1
  {
    \tag_if_active:T
      {
        \seq_if_empty:NF \g__tbl_struct_cur_seq
          { \tagpdfsetup {table/multirow=#1} }
      }
  }
\NewDocumentCommand \AccessibleInlineMath { m m }
  {
    \group_begin:
      \tagpdfsetup {math/alt/use=true}
      \tl_set:Nn \l__math_content_template_tl {#1}
      $#2$
    \group_end:
  }
\ExplSyntaxOff
\makeatother
\makeatletter
\AddToHook{begindocument/end}[acmart-arxiv-tagging/artifact-rules]{%
  \let\ACMtag@originalVline\vline
  \def\vline{\ACMtag@artifactBegin\ACMtag@originalVline\ACMtag@artifactEnd}%
  \patchcmd{\@BTnormal}
    {\hrule\@height\@thisrulewidth}
    {\ACMtag@artifactBegin\hrule\@height\@thisrulewidth\ACMtag@artifactEnd}
    {}{\PackageWarning{acmart-arxiv-tagging}{Booktabs Artifact rule patch did not match}}%
  \@ifundefined{CT@@do@color}{}{%
    \let\ACMtag@originalCellColor\CT@@do@color
    \def\CT@@do@color{\ACMtag@artifactBegin\ACMtag@originalCellColor\ACMtag@artifactEnd}%
  }%
  \let\ACMtag@originalFootnoteRule\footnoterule
  \def\footnoterule{\ACMtag@artifactBegin\ACMtag@originalFootnoteRule\ACMtag@artifactEnd}%
  \let\ACMtag@originalCopyrightRule\copyrightpermissionfootnoterule
  \def\copyrightpermissionfootnoterule{\ACMtag@artifactBegin\ACMtag@originalCopyrightRule\ACMtag@artifactEnd}%
  \@ifundefined{defaultfootnoterule}{}{%
    \let\ACMtag@originalDefaultFootnoteRule\defaultfootnoterule
    \def\defaultfootnoterule{\ACMtag@artifactBegin\ACMtag@originalDefaultFootnoteRule\ACMtag@artifactEnd}%
  }%
}
\makeatother

\makeatletter
\AddToHook{begindocument/before}[acmart-arxiv-tagging/backtick-map]{%
\immediate\pdfobj stream {/CIDInit /ProcSet findresource begin
12 dict begin
begincmap
/CIDSystemInfo << /Registry (Adobe) /Ordering (UCS) /Supplement 0 >> def
/CMapName /ACMBacktick-UCS def
/CMapType 2 def
1 begincodespacerange
<00> <FF>
endcodespacerange
1 beginbfchar
<60> <0060>
endbfchar
endcmap
CMapName currentdict /CMap defineresource pop
end
end}
\edef\ACMBacktick@cmapref{\the\pdflastobj\space 0 R}
\DeclareTextCommand{\textasciigrave}{TS1}{%
  \ifnum\pdfmatch{^tctt0800}{\fontname\font}>0
    \pdffontattr\font{/ToUnicode \ACMBacktick@cmapref}%
  \fi
  \char96\relax
}
}%
\makeatother

\ExplSyntaxOn
\tl_new:N \l__acmtag_table_summary_tl
\tl_new:N \l__acmtag_table_summary_pdf_tl
\cs_new_protected:Npn \__acmtag_table_add_summary:
  {
    \tag_if_active:T
      {
        \tl_if_empty:NF \l__acmtag_table_summary_tl
          {
            \pdfstringdef \l__acmtag_table_summary_pdf_tl
              {\tl_use:N \l__acmtag_table_summary_tl}
            \tag_struct_gput:ene {\g__tbl_struct_table_tl} {attribute}
              {/O /Table /Summary (\l__acmtag_table_summary_pdf_tl)}
          }
      }
  }
\NewTaggingSocketPlug{tbl/hmode/begin}{ACMTableSummary}
  {
    \use:c {__socket_tagsupport/tbl/hmode/begin_plug_Table:w}
    \__acmtag_table_add_summary:
  }
\NewTaggingSocketPlug{tbl/vmode/begin}{ACMTableSummary}
  {
    \use:c {__socket_tagsupport/tbl/vmode/begin_plug_Table:w}
    \__acmtag_table_add_summary:
  }
\NewDocumentCommand \AccessibleTableDescription {m}
  {
    \tl_set:Nn \l__acmtag_table_summary_tl {#1}
    \AssignTaggingSocketPlug{tbl/hmode/begin}{ACMTableSummary}
    \AssignTaggingSocketPlug{tbl/vmode/begin}{ACMTableSummary}
  }
\ExplSyntaxOff

\documentclass[manuscript,screen,nonacm]{acmart}
\hypersetup{pdfdisplaydoctitle=true}
\usepackage{colortbl}
\usepackage{tabularx}
\usepackage{listings}

\newcommand{\name}{\textit{AssemblyLens}}
\AtBeginDocument{%
  }

\setcopyright{acmlicensed}
\copyrightyear{2027}
\acmYear{2027}
\acmDOI{XXXXXXX.XXXXXXX}

\acmConference[CHI '27]{CHI Conference on Human Factors in Computing
  Systems}{May 10--14, 2027}{Pittsburgh, PA, USA}
\acmISBN{978-1-4503-XXXX-X/27/05}

\tagpdfsetup{table/header-rows={1}}
\begin{document}

%%
%% The "title" command has an optional parameter,
%% allowing the author to define a "short title" to be used in page headers.
\title{AssemblyLens: Exploring the Role of Structural Roadmaps and Strategies of Furniture Assembly Used by Blind and Low Vision Individuals}

% AssemblyLens: exploring mental models and strategies during Blind and Low Vision individuals furniture assembly

%%
%% The "author" command and its associated commands are used to define
%% the authors and their affiliations.
%% Of note is the shared affiliation of the first two authors, and the
%% "authornote" and "authornotemark" commands
%% used to denote shared contribution to the research.
\author{Bo Liu}
\affiliation{
  \institution{Cornell Tech}
  \city{New York, New York}
  \country{USA}}
\email{Bl685@cornell.edu}

\author{Gene S-H Kim}
\affiliation{
  \institution{Massachusetts Institute of Technology}
  \city{Cambridge, Massachusetts}
  \country{USA}}

\author{Franklin Mingzhe Li}
\affiliation{
  \institution{UNC-Chapel Hill}
  \city{Chapel Hill, North Carolina}
  \country{USA}}

\author{Rajalakshmi Nandakumar}
\affiliation{%
  \institution{Cornell Tech}
  \city{New York, New York}
  \country{USA}}

\author{Thijs Roumen}
\affiliation{%
  \institution{Cornell Tech}
  \city{New York, New York}
  \country{USA}}
\email{thijs.roumen@cornell.edu}
%%
%% By default, the full list of authors will be used in the page
%% headers. Often, this list is too long, and will overlap
%% other information printed in the page headers. This command allows
%% the author to define a more concise list
%% of authors' names for this purpose.
\renewcommand{\shortauthors}{Trovato et al.}

%%
%% The abstract is a short summary of the work to be presented in the
%% article.
\begin{abstract}
Instructions to assemble flat-packed furniture rely on visuals, making them inaccessible to Blind and Low Vision (BLV) individuals. Consequently, BLV individuals depend on caregivers or external services (e.g., \textit{Be My Eyes}), limiting their autonomy. We explore how to empower them to assemble such furniture themselves, by providing eight BLV participants in our study with accessible instructions consisting of a high-level roadmap and stepwise procedural guidance. We use a technology probe, \name{}, which converts visual assembly instructions into screen-reader-accessible text. We derive insights into converting instructions, empirical observations of BLV assembly strategies, and the role the roadmap played in them. Our observations yield three themes: spatial cognition and mental mapping, embodied ergonomics and kinetic semantics, and assembly workflow and user agency. We present eight design recommendations for instruction system design. Our findings are a step toward an agenda for inclusive instruction in physical tasks.

\end{abstract}

%%
%% The code below is generated by the tool at http://dl.acm.org/ccs.cfm.
%% Please copy and paste the code instead of the example below.
%%


% \ccsdesc[500]{Do Not Use This Code~Generate the Correct Terms for Your Paper}
% \ccsdesc[300]{Do Not Use This Code~Generate the Correct Terms for Your Paper}
% \ccsdesc{Do Not Use This Code~Generate the Correct Terms for Your Paper}
% \ccsdesc[100]{Do Not Use This Code~Generate the Correct Terms for Your Paper}

%%
%% Keywords. The author(s) should pick words that accurately describe
%% the work being presented. Separate the keywords with commas.
%% A "teaser" image appears between the author and affiliation
%% information and the body of the document, and typically spans the
%% page.

\begin{teaserfigure}
  \includegraphics[alt={Three photographs and a workflow trace show P1 assembling a shoe rack. The photographs show P1 handling a partial frame, working on its connections, and installing a fabric shelf. The trace follows five events: starting to listen, skipping the structural roadmap, disassembling, returning to the roadmap, and finishing assembly. Its backward and upward movements show revisiting instructions and undoing assembly before progressing to completion.},width=\textwidth]{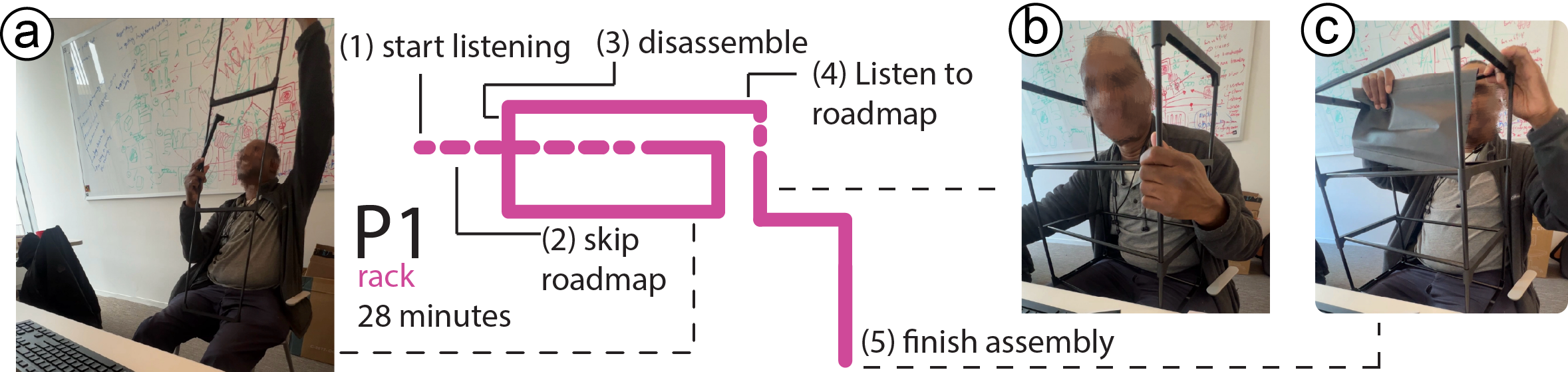}
  \caption{A participant’s shoe rack assembly workflow using \name{} instructions. Horizontal movement represents progress through instructions read aloud by a screen reader; vertical movement represents physical assembly progress. (a)~The participant skips the structural roadmap and starts assembling. (b)~They backtrack, disassemble the frame, and revisit the roadmap. (c)~They follow the procedural instructions to complete the assembly task. Total task time was 28~minutes.}
  \Description{Three photographs and a workflow trace show P1 assembling a shoe rack. The photographs show P1 handling a partial frame, working on its connections, and installing a fabric shelf. The trace follows five events: starting to listen, skipping the structural roadmap, disassembling, returning to the roadmap, and finishing assembly. Its backward and upward movements show revisiting instructions and undoing assembly before progressing to completion.}
  \label{fig:teaser}
\end{teaserfigure}

%%
%% This command processes the author and affiliation and title
%% information and builds the first part of the formatted document.
\maketitle

\section{Introduction}
Assembling flat-packed furniture is challenging, particularly when it must be completed non-visually~\cite{Thosewho73:online}. Assembly manuals commonly use images to communicate parts, spatial relationships, and assembly steps. This visual format creates accessibility barriers for Blind and Low Vision (BLV) individuals. Additionally, there is no design guidance on how to make assembly instructions accessible for BLV individuals. Even with recent developments such as Vision-Language Models (VLMs), simply transcribing graphic manuals into text-based instructions is insufficient and possibly misleading~\cite{reddy2026lost}. We argue that truly empowering BLV individuals requires a different way of thinking rather than simple translation: it may preserve the procedural steps but lacks the overview required to understand how parts align to form a bigger picture~\cite{daniel2012put}.

Prior research has explored techniques to reduce dependence on visual instruction. Examples include techniques such as embossed braille~\cite{chang2023understanding}, embroidered tactile graphics~\cite{seehorn2025beyond}, tactile aids for laser cut assembly~\cite{chang2021daedalus}, and auditory markers~\cite{shi_tickers_2016} placed directly on objects. While effective, such approaches require modifying the furniture itself, hindering broad adoption. Consequently, BLV individuals depend on sighted assistance, either in person or via services like Be My Eyes\footnote{www.bemyeyes.com} and Aira\footnote{https://aira.io/}. Relying on such assistance limits autonomy, and AI-based alternatives such as “Be My AI”\footnote{https://www.bemyeyes.com/bme-ai/} leave users unable to verify what they are told. This friction frequently leads BLV individuals to avoid flat-packed products altogether~\cite{reddy2026lost}, opting instead for pricier pre-assembled furniture, external assembly services, or help from family.

Achieving independent eyes-free assembly requires more than translating visual instruction; it requires an understanding of the unique BLV perspective. Similar to how a sighted user might first browse through the manual before assembling, understanding the structure first helps to know \textit{why} each instruction makes sense. In line with spatial cognition theory, assembly requires both route knowledge (procedural actions) and “survey knowledge” (a structural roadmap)~\cite{daniel2012put, siegel1975development, thorndyke1982differences}. Without such survey knowledge, BLV individuals may assemble the furniture incrementally from instructions, placing demands on users' memory while manipulating physical parts (a demanding action in itself). This makes it hard, if not impossible, to verify steps or backtrack when making mistakes~\cite{loomis2012representing, reddy2026lost}. To develop an understanding of what it entails to create accessible assembly instruction, we ask the following research questions:

\begin{itemize}
    \item \textbf{RQ1:} How can we develop instructions to support BLV individuals in developing survey knowledge as well as procedural guidance for furniture assembly, based on existing visual instructions?
    \item \textbf{RQ2:} What physical assembly strategies do BLV individuals use when following procedural assembly instructions, and what role does a roadmap play in providing structural survey knowledge during later assembly steps?
\end{itemize}

A key challenge for RQ1 is to develop language to describe the structural roadmap of the assembly through audio. To this end we conducted an informal formative study to identify the information necessary for BLV individuals to build an accurate mental model of furniture assembly. We first created a highly verbose structural roadmap, which we incrementally simplified, while asking sighted pilot participants to draw the furniture we described, and our BLV co-author to reproduce it in clay. We identified design principles indicating what language worked well and what to avoid when conveying survey knowledge of assembly. 

To understand how such survey knowledge helps BLV individuals during assembly, we developed a technology probe we call \name{}, which follows the identified design principles to convert existing visual assembly instruction into text instructions comprising a structural roadmap and step-by-step procedural guidance. By providing concrete instructions as an example, we elicit responses from BLV individuals and observe their physical assembly strategies. No set of instructions is perfect, and it is precisely participants' reactions to a concrete, imperfect one that we consider valuable for future development. Developing a technology probe is the preferred study methodology to explore this question.

In response to RQ2, we conducted an in-depth evaluation with eight BLV participants assembling two distinct pieces of furniture to understand how to advance instruction for physical tasks. Through our study, we found that: 1.~Roadmaps allow users to build a mental “image” that supports the later assembly process. 2.~We highlight how users rely on tactile feedback during procedural execution. 3.~We identify underexplored research areas, such as embodied ergonomics to assist alignment and verification. Through this exploration, we lay out concrete design guidelines to improve future non-visual assembly instructions.

\section{Related Work}
Our work builds on prior research in accessible support for spatial tasks, the design of instruction manuals, computer-guided assembly, and generative AI for non-visual task guidance. 

\subsection{Accessibility in Spatial Tasks}
Prior work has explored how BLV individuals perform spatial manipulation tasks using tactile materials~\cite{seehorn2025beyond,phutane2022tactile} and 3D printed models~\cite{siu2019shapecad,kim2015toward,nagassa2025enhancing}, showing how tactile representations can make otherwise visual structures more accessible. Research on non-visual cooking has examined sensory aspects of food choice~\cite{kostyra2017food} and how cooks obtain contextual information about kitchen objects~\cite{li2024contextual}. Through interviews and recipe-review activities, \citet{li2024recipe} found that participants wanted an overview at the beginning of a recipe. Participants also requested brief summaries of the main action and estimated duration at the beginning of individual steps to make navigation easier. Vid2Coach~\cite{huh2025vid2coach} transforms how-to videos into accessible task assistants that provide instructions and context-aware feedback. Complementing instruction delivery, OSCAR~\cite{li2025exploring} explores object-status recognition to track progress through cooking recipes.

Related work has also examined non-visual support for understanding 3D structures and assembly. \textit{TouchPilot}~\cite{wang2023touchpilot} and \textit{Touching Movement}~\cite{tanaka2026touching} support BLV individuals in recognizing 3D structures through guided exploration, while \citet{matsumoto2025assembling} show that successful non-visual assembly support requires more than simple part identification alone. Effective support must also communicate spatial orientation, procedural progress, and the relationships among parts so that users can maintain an accurate understanding of the evolving structure throughout the task. These findings motivate our attention to how instructions communicate component relationships and assembly actions.

\subsection{Designing Instruction Manuals}

Effective assembly guidance draws from long-standing principles of instruction design. \citet{novick2000folding} showed that both procedural diagrams and final-state diagrams play an important role in supporting assembly performance. \citet{agrawala2003designing} further argued that effective assembly instructions depend on both planning and presentation, and derived visual design principles that communicate grouping, sequence, orientation, and part relationships clearly. Beyond diagrams, \citet{eiriksdottir2011procedural} reviewed the roles of procedural steps, explanatory principles, and examples, emphasizing that their effectiveness depends on whether the goal is immediate task performance, learning, or transfer. 

Research on assembly instruction design also addresses how guidance should be organized. \citet{hinrichsen2022design} recommend providing an initial overview image of the target product, required parts, and assembly operations. \citet{bartolomei2025streamlining} introduced S-AID, a systematic framework for decomposing assembly tasks, identifying information requirements, and selecting presentation formats.

Automated generation offers another way to develop assembly guidance. \citet{maeda2025additional} used GPT-4 to generate captions accompanying furniture assembly instructions. Their comprehension-task evaluation found that erroneous captions reduced response accuracy, illustrating the importance of checking generated descriptions. In the context of accessibility, “Lost in Instructions”~\cite{reddy2026lost} examines how BLV individuals use product manuals together with AI-rewritten instructions for assembly, operation, and troubleshooting tasks. Their findings show that manuals remain essential resources, while AI-generated guidance can remain incomplete, incoherent, or misleading.

Building on this work, we investigate how graphical furniture manuals can be translated into a structural roadmap describing major components and their spatial relationships, followed by procedural instructions.

\subsection{Computer-Guided Assembly}
Research in computer vision and robotics has developed methods for inferring assembly structure and planning physical actions. IKEABot~\cite{knepper2013ikeabot} coordinates multiple robots using assembly plans inferred from geometric specifications. More recently, Manual-PA~\cite{zhang2025manual} uses instruction diagrams to predict assembly order and 6D part poses. Manual2Skill~\cite{tie2025manual2skill} extracts hierarchical assembly graphs from manual images and combines pose estimation with motion planning for robotic assembly.

For human-robot collaborative assembly, \citet{wang2024assembly} construct a graph of part connections from 2D graphical instruction manuals and allocate tasks based on assembly stability and complexity. Their approach also incorporates actions needed to support parts, reorient subassemblies, and change tools. Related dataset resources include FurnitureBench~\cite{heo2025furniturebench}, which provides standardized robotic assembly tasks using 3D-printed furniture, and IKEA-Manual~\cite{wang2022ikea}, which pairs graphical manuals with annotated 3D part models.

Together, these approaches show how assembly information can be represented computationally to support planning and execution.

\subsection{Generative AI for Non-Visual Guidance of Spatial Tasks}

Generative AI provides opportunities to support non-visual guidance. Scene-description technologies have made it easier to inspect photographs, identify objects, and ask follow-up questions about visual scenes~\cite{gonzalez2024investigating}. This includes commercial apps, such as “Be My AI”\footnote{\url{https://www.bemyeyes.com/bme-ai/}}, which enable rich, conversational interactions with a user's visual surroundings. Broadening this scope, studies show that blind users are also using general-purpose generative AI tools, such as ChatGPT, for tasks like document summarization, knowledge access, and daily problem-solving~\cite{adnin2024look, tang2025everyday}. 

At the same time, prior work consistently identifies limitations around user trust, algorithmic hallucination and incomplete semantic grounding~\cite{10.1145/3746059.3747597, gonzalez2025towards, xie2025beyond}. To reduce the burden on users to verify or contest AI outputs, researchers have explored interaction techniques that help users inspect and evaluate generated content. For example, \textit{GenAssist} utilizes visual questions to help creators verify generated images~\cite{huh2023genassist}. \textit{A11yShape} links code, component hierarchies, descriptions, and rendered geometry to support non-visual inspection and editing of 3D models~\cite{zhang2025a11yshape}. 

Together, these studies motivate examining both the generation of non-visual instructions and their use during physical tasks. We use \name{} to investigate how BLV individuals interpret and act on structural roadmaps and procedural instructions derived from graphical furniture manuals. Our study examines how this guidance interacts with tactile exploration and assembly strategies, informing design recommendations for future non-visual instructions.

\section{Informal Formative Study: Communicating Survey Knowledge}

Consistent with prior work on task previews for BLV users~\cite{huh2025vid2coach}, we provide a structural roadmap before the procedural instructions. This roadmap describes the furniture's major components, their spatial relationships, and how they connect to form the assembled object, with the aim of supporting users' development of survey knowledge. We conducted an informal formative study to explore how the choice of words and level of detail could communicate this structure clearly while keeping the roadmap concise.

\subsection{Process}
% talking how we made piplines for later VLM
During the exploration, we used the IKEA Stefan Chair\footnote{\url{https://www.ikea.com/us/en/p/stefan-chair-brown-black-00211088/}} as a representative piece of furniture. Initially, we designed structural descriptions featuring exhaustive details about the look and shape of every piece, as shown in Appendix ~\ref{appendix:initial_instructions}. Our goal was to systematically reduce this linguistic load to its essential information.

We conducted pilot studies with four sighted participants who drew the furniture based on the auditory description alone. Each participant heard a single version of the description, delivered by a researcher reading aloud, and then drew the chair without being able to ask clarifying questions. After each pilot, we refined the language to identify critical information gaps and spatial ambiguities, iterating until participant drawings reflected the furniture structures using a minimal description.

\subsection{Design Principles}

After four iterations, and one with our BLV co-author, we derived the following principles to guide our computational translation: 

\subsubsection{Selective Information Inclusion}
Not all visual information is equally relevant, and excessive information is cognitively challenging. We determine inclusion in the roadmap based on:
\begin{itemize}
\item Prioritize parts that can be identified by a name (e.g., “seat”), if not, they  require additional description based on tactile properties(e.g., “the seat is a round shape flat plate”).
\item Describe how major parts physically connect and their structural purpose (e.g., “the rail connects the two H-frames”).
\item Exclude explanations of minor fasteners, such as individual bolts, screws, or nuts, to maintain focus on the main components.
\end{itemize}

\subsubsection{Choice of Words}
We found that it is important to carefully weigh how to exploit analogies and metrics and which ones to use.
\begin{itemize}
\item Use relative words: When describing actions or selection, provide comparative (e.g., “wider side”) and force-related words (e.g., “firmly press”) to guide selection and exploration.
\item Prefer relative bodily or spatial anchors over exact measurements: avoid using metrics like centimeters or inches. Instead, use relative structural anchors (e.g., “near the floor”).
\item Avoid generic labels (e.g., “Part A”, “Part 1”). Instead, use tactile properties.
\end{itemize}

\subsubsection{Minimizing Cognitive Load}
The sequential nature of audio requires users to build a working memory for everything that was said. To prevent users from losing patience or forgetting details, we apply the following rules:
\begin{itemize}
\item Replace exhaustive descriptions with common knowledge (e.g., “a standard rectangular backrest”) when precise dimensions are not critical for assembly.
\item Minimize the information provided before a user physically interacts with a part to reduce the burden on short-term memory. Informed by our formative observations, we adopted approximately 300 words as a practical guideline to keep the roadmap concise. 
\end{itemize}

\subsection{Validation} 
\begin{figure}
    \centering
    \includegraphics[alt={Reference furniture images are paired with clay representations created from structural roadmap descriptions. The top row shows a three-level open nightstand and two views of its clay representation. The bottom row shows a shoe rack and its clay representation. The clay models use horizontal surfaces and upright supports to represent the furniture's overall structure and the relationships between its levels.},width=0.5\linewidth]{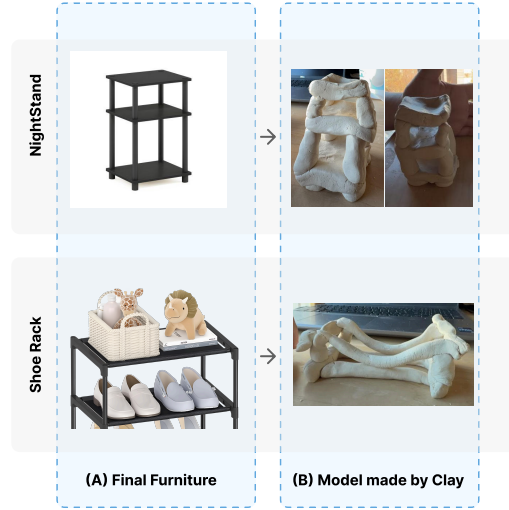}
\Description{Reference furniture images are paired with clay representations created from structural roadmap descriptions. The top row shows a three-level open nightstand and two views of its clay representation. The bottom row shows a shoe rack and its clay representation. The clay models use horizontal surfaces and upright supports to represent the furniture's overall structure and the relationships between its levels.}
    \caption{The two pieces of furniture a BLV co-author built with clay. (a)~ the furniture final images (b)~ Model made by clay, based on structural roadmap descriptions, structurally resembling the target assembly.} 
    \label{fig:formative-clay}
\end{figure}

To validate the principles, we generated roadmaps for two furniture pieces previously unknown to our BLV co-author. As shown in Figure~\ref{fig:formative-clay}, the co-author constructed representations using clay based on these instructions, as a low-fidelity technique to verify their understanding of the furniture. 

This is an informal formative study with the BLV co-author; the description style is evaluated in our main study with eight BLV participants. Its purpose here is to establish that the principles were worth building on before we committed them to the pipeline. They come from a single piece of furniture, so we expect them to need adjustment for assemblies unlike it. They were sufficient to produce every roadmap used in the rest of this paper, and we offer them as a concrete starting point for others building non-visual instructions but caution against using them as generalized principles.

\section{\name{}: A Probe to Study AI Generated Assembly Instruction}
% talking how we designed interface and interaction guidance. 
To study how participants respond to survey knowledge and how they execute physical assembly instructions, we developed a tech probe \name{}, which converts existing graphical manuals into accessible text instructions that can be read aloud by screen readers. To assess whether AI generated conversion is realistic, we opted to build a probe/tool rather than using hand-written instruction. In this section, we lay out the pipeline for the probe and provide a short technical evaluation to get a sense of its feasibility. All instructions used in the study are generated by this probe. The primary users of the probe are thus the research team to convert existing manuals. The exported instructions are directly used by BLV individuals in our study without further modification.

\subsection{System Pipeline}
\begin{figure}
    \centering
    \includegraphics[alt={A left-to-right flowchart shows the AssemblyLens pipeline and human review at each of its three stages. An assembly manual enters Step 1, which produces a catalog of extracted parts with numbered bounding boxes. Step 2 links extracted parts to locations in assembly diagrams. Step 3 produces instruction text, followed by an accessible-text output. Beneath each stage, a review panel connects back to that stage: reviewers can add or resize part boxes, move or clear connection markers, and edit and save instructions. Purple arrows connect the processing stages; teal arrows connect each stage to its review and confirmation controls.},width=1\linewidth]{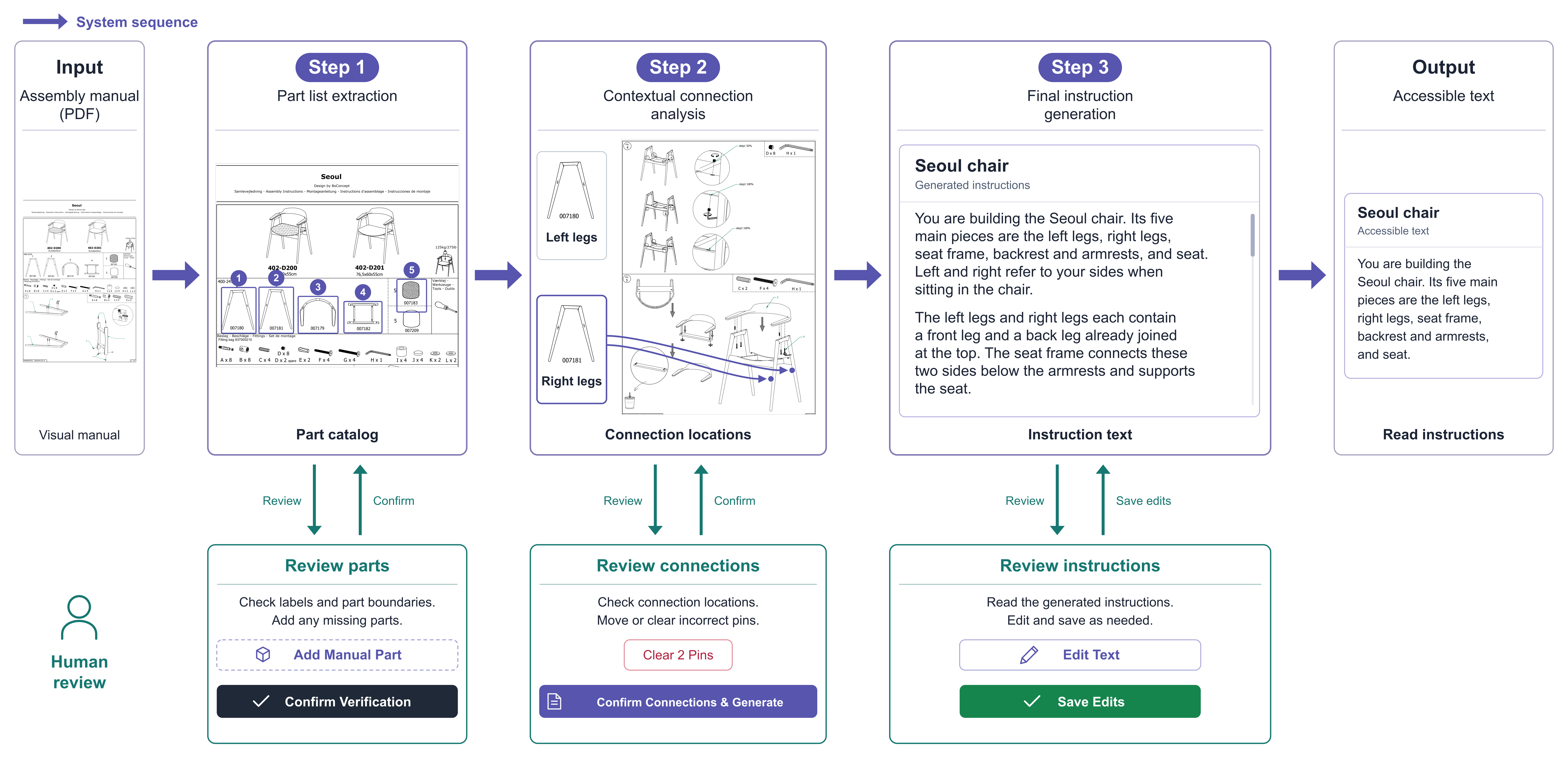}
\Description{A left-to-right flowchart shows the AssemblyLens pipeline and human review at each of its three stages. An assembly manual enters Step 1, which produces a catalog of extracted parts with numbered bounding boxes. Step 2 links extracted parts to locations in assembly diagrams. Step 3 produces instruction text, followed by an accessible-text output. Beneath each stage, a review panel connects back to that stage: reviewers can add or resize part boxes, move or clear connection markers, and edit and save instructions. Purple arrows connect the processing stages; teal arrows connect each stage to its review and confirmation controls.}
    \caption{The \name{} pipeline.
    Given a PDF assembly manual, the system (1) extracts structural
    parts, (2) maps parts to their locations in the assembly diagrams
    and identifies their functional roles, and (3) generates a
    structural roadmap and procedural instructions. Human reviewers
    verify and correct the part detections, connection locations,
    and generated text at the corresponding stages. Purple arrows
    indicate the system sequence, and teal arrows indicate human
    review and correction. The final output is accessible text
    that can be read aloud using screen readers.}
    \label{fig:system_steps}
\end{figure}

The pipeline is driven by a multi-modal large language model (MLLM) (Gemini 3 Pro~\cite{gemini3report2025}). It takes the PDF of an assembly manual as input, and converts it to text instruction using three distinct steps: (1)~part extraction, (2)~connection analysis, and (3)~final instruction generation. JSON strings serve as the interchange format between these steps, which allows for easy editing if errors accumulate in each step. Fig.~\ref{fig:system_steps} shows the pipeline of \name{}.

\subsubsection{Step 1: Part Extraction}
\name{} gets invoked by loading an instruction manual (currently supporting PDF) into the web interface. \name{} identifies structural parts (e.g., seat, armrest, backrest) while intentionally filtering out small hardware (e.g., fasteners, washers). 
It draws bounding boxes around each structural part within the manual, to let users verify whether all parts are correctly captured. Visual instructions typically show parts from different perspectives. \name{} prompts the MLLM to first store all parts with their bounding box and then de-duplicates them. Each selected component is captured as:

\AccessibleInlineMath{Record fields: name, label, page index, and two dimensional bounding box.}{\{\texttt{name, label, pageIndex, box\_2d}\}}

These are presented to users to allow for human verification, before moving on with the next steps. For example, users can resize bounding boxes and add or remove components. Upon confirmation, \name{} generates an image that includes all parts, each with its name, either captured from the guide or generated by the model based on the parts' shapes. See Fig.~\ref{fig:system_steps} step 1. 
% How we expand our guidance/ instruction into all prompts and how our current system works. 

\subsubsection{Step 2: Connection Analysis}
\name{} uses the part list of step 1 to identify connections, and constructs an assembly graph between part image and representation in visual instructions~\cite{tie2025manual2skill,tie2025manual2skill++,liu2024grounding}. 

To determine the functional role of each component, we apply a Chain-of-Thought~\cite{wei2022chain} prompting strategy to MLLM. Rather than directly predicting the role of each component, our MLLM is prompted to first locate each part's spatial reference within visual assembly instruction where the part is visually identifiable. Using this localization step as an additional context can ground model inference in visual evidence, which helps to reduce hallucinated component. The connection diagram is shown to users to review, shown as Fig.~\ref{fig:system_steps} Step 2. Given the part images and reference from visual manuals, we then prompt the model to generate a functional explanation for each part (full prompt can be found in appendix~\ref{appendix:prompts}). 

To verify this functional explanation, the MLLM is also required to generate a text justification explaining its reasoning based on the instruction reference and the component's shape. If the MLLM cannot locate a part due to visual occlusions or ambiguous perspective drawings in the manual, it assigns a "low" confidence score to its prediction. \name{} flags any parts with "low" confidence score in connection diagram for user verification. Through the interface, the user can review the relation between the isolated part image and the model's predicted target location, and adjust the pointing location as needed. If the user updates the target area, \name{} re-executes this contextual analysis based on the corrected spatial constraints and updates the component's functional explanation for all subsequent steps. Each labeled part now contains a brief context explanation for the roles in the final assembly:

\AccessibleInlineMath{Record fields: name, label, and explanation.}{\{\texttt{name, label, explanation}\}}

\subsubsection{Step 3: Final Instruction Generation}
Based on the part list and assembly graph, \name{} prompts the MLLM to generate the roadmap and procedural instructions. The prompt contains the principles of our formative study as a context. The final result is exported as a text document which can be authored through \name{}. By exporting instructions as plain text, our system allows the guidance to be presented as speech or on refreshable braille displays through compatible screen readers.

\subsection{Technical Evaluation}
To assess technical feasibility on real assembly manuals, we created a benchmark and conducted a technical evaluation of resulting assembly instructions. 

\subsubsection{Composing the Benchmark}

We included manuals that provided a parts list and were publicly available as PDFs. We sourced 34 assembly manuals across 25 furniture suppliers (the full list is available in appendix~\ref{apx:benchmark}). Table ~\ref{tab:representative_models} provides an overview of the type of models and instructions included in this benchmark.

\begin{table*}[t]
\AccessibleTableDescription{The benchmark contains 34 furniture assembly manuals from 25 manufacturers, presented in three pairs of furniture-type and manufacturer columns. Furniture includes chairs, benches, a bar stool, tables, storage furniture, carts, and floor lamps. BoConcept appears in six entries, Furniture of America in four, CIOC in two, and the remaining 22 manufacturers in one entry each.}
\centering
\resizebox{\textwidth}{!}{%
\begin{tabular}{ll|ll|ll}
\toprule
\textbf{Type} & \textbf{Manufacturer} & \textbf{Type} & \textbf{Manufacturer} & \textbf{Type} & \textbf{Manufacturer} \\
\midrule
Armchair & Yaheetech Store & Chair & CLATINA & Ottoman & Furniture of America \\
Bar Stool & BoConcept & Chair & Kyajah & Shelf Unit & Furniture of America \\
Bench & BoConcept & Desk Chair & Sweetcrispy & Shoe Rack & Autonomier \\
Bench & Furniture of America & Dining Chair & BoConcept & Side Table & GKAN \\
Bookcase & nathanjames & Floor Lamp & CIOC & Side Table & Highland Dunes \\
Bookcase & VASAGLE & Floor Lamp & CIOC & Sideboard & RoyalCraft \\
Bookshelf & NicBex & Hanging Chair & Inspire & Storage Bench & KaBuea \\
Cabinet & DSHADE & Nightstand & Furinno & Table & BoConcept \\
Cart & Furniture of America & Office Chair & BOWTHY & Table & DELLMAO \\
Cart & O\&K & Office Chair & TECHNI MOBIL & Table & Modway \\
Chair & BoConcept & Office Chair & Zesthouse & & \\
Chair & BoConcept & Ottoman & DREAMS IDEA & & \\
\bottomrule
\end{tabular}%
}
\caption{The 34 models of the \name{} benchmark. Full list with more details is included in appendix~\ref{apx:benchmark}.}
\label{tab:representative_models}
\end{table*}

% \begin{figure}
%     \centering
%     \includegraphics[width=1\linewidth]{figures/fig-tmp-1col.jpg}
%     \caption{Some models of the \name{} benchmark and their instructions. Full list is included in appendix~\ref{apx:benchmark}.}
%     \label{fig:representative_models}
% \end{figure}

\subsubsection{Phase-by-phase Evaluation}
We independently assessed the success of Step 1 (Part Extraction), Step 2 (Connection Analysis), and Step 3 (Final Instruction Generation). We measured the success rate of \name{}'s automated conversion pipeline and quantified the number of user interactions in the human verification interface to correct identified errors.

\subsubsection{Step 1: Part Extraction Achieves 94.1\% Success Correction Requires Three Steps of Manual Adjustment}
We assessed \name{}'s ability to extract parts and generate accurate bounding boxes.

\textbf{Success Criteria:} We define a successful extraction when the model extracts all parts without missing or cutting out or overlapping with other components.

\textbf{Technical Results:} Across 34 manuals containing 296 expected structural-part detections, \name{} correctly extracted 294 parts automatically (99.3\%). Human intervention was required for 2 manuals (5.9\%); 32/34 manuals (94.1\%) required no correction at this stage.

\begin{figure}[b]
    \centering
    \includegraphics[alt={Two examples show structural-part extraction errors and their correction. Each panel combines a source-manual screenshot with editing-interface screenshots. In A, one bounding box encloses two distinct leg frames; a reviewer adds another box and adjusts the boxes to represent the parts separately. In B, a misplaced bounding box fails to enclose the intended component drawing; the reviewer moves and enlarges the box to enclose it.},width=1\linewidth]{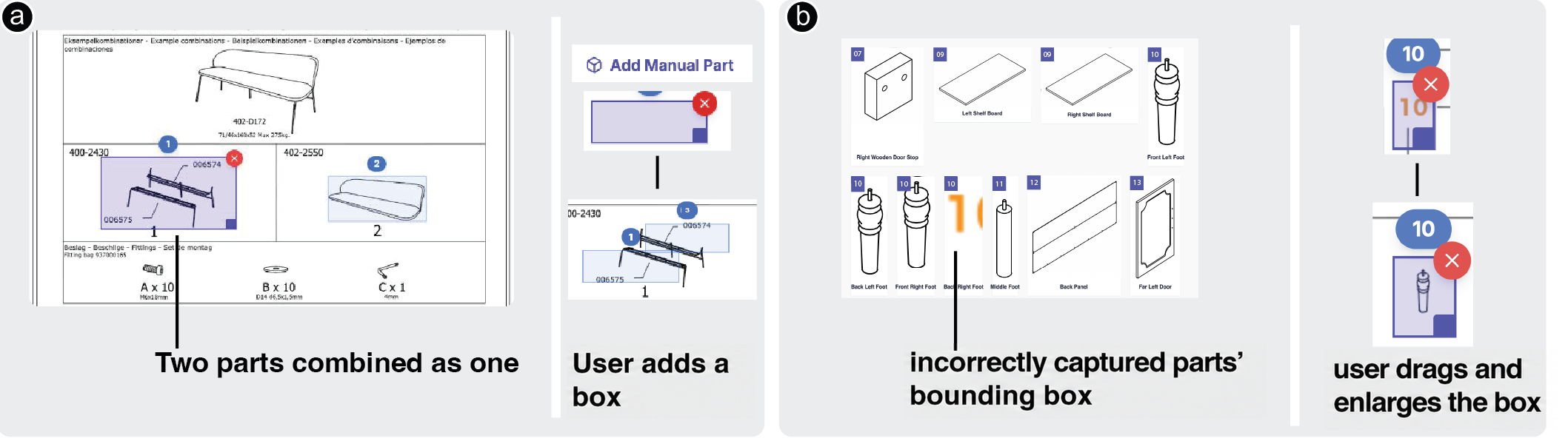}
\Description{Two examples show structural-part extraction errors and their correction. Each panel combines a source-manual screenshot with editing-interface screenshots. In A, one bounding box encloses two distinct leg frames; a reviewer adds another box and adjusts the boxes to represent the parts separately. In B, a misplaced bounding box fails to enclose the intended component drawing; the reviewer moves and enlarges the box to enclose it.}
    \caption{Failure modes during the automated processing stages of \name{}. A and B illustrate localization challenges in Step 1 (Part Extraction).}
    \label{fig:error_analysis_part1}
\end{figure}

\textbf{Human intervention:} 
Specifically, the manual in Figure \ref{fig:error_analysis_part1} groups two distinct “leg frames” under a single label without providing the visual cues necessary for automatic segmentation. This case requires the user to click “Add Manual Part” and drag the created bounding box to the target area. Users also adjust the previous bounding box that covered two components to cover only one component (three steps in total). Another manual utilized a single part image to represent multiple leg images, but one part was incorrectly captured. This requires the user to drag the current bounding box to the target leg part in the manual, and enlarge the bounding box size to cover the part (two steps in total).

\subsubsection{Step 2: Connection Analysis Automatically Achieved 88.2\% Success, Correction Requires up to 2 Manual Interventions}
The second step constructs the structural connection between parts and visual instructions. 

\textbf{Success Criteria:} We manually checked whether the arrows linking extracted parts to their locations in the assembly diagrams pointed to the correct components. A manual passed this stage if no incorrect or missing links were identified. 

\textbf{Technical Results:} The system automatically processed 30/34 manuals successfully and with human intervention all manuals were successful. We identify two types of errors requiring human intervention. Three cases contained incorrect links and two cases contained missing links; one model contained both of these cases.

\begin{figure}
    \centering
    \includegraphics[alt={Two examples show incorrect mappings between extracted parts and assembly diagrams. Arrows connect isolated part images to predicted locations, with enlarged views showing incorrect and corrected selections. In A, a wooden brace is mapped to the wrong location in the assembly. In B, a long connector is mapped to empty space. In both cases, a reviewer replaces the incorrect target with the component's correct location.},width=1\linewidth]{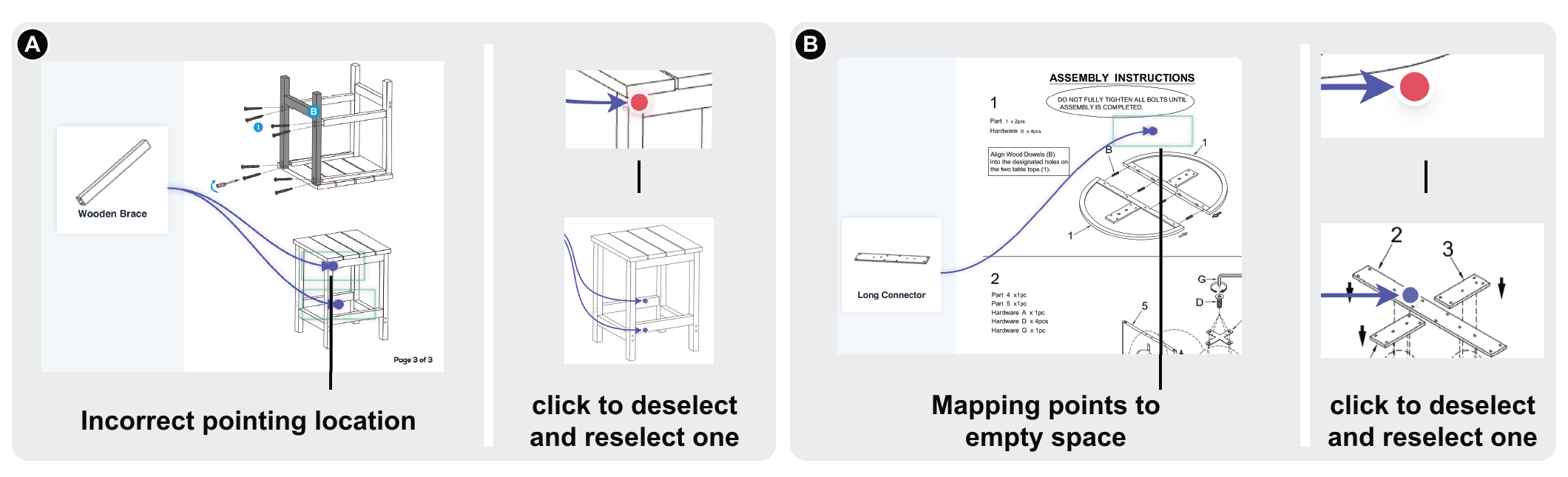}
\Description{Two examples show incorrect mappings between extracted parts and assembly diagrams. Arrows connect isolated part images to predicted locations, with enlarged views showing incorrect and corrected selections. In A, a wooden brace is mapped to the wrong location in the assembly. In B, a long connector is mapped to empty space. In both cases, a reviewer replaces the incorrect target with the component's correct location.}
    \caption{Failure modes during the automated processing stages of \name{}. We demonstrate semantic mapping errors in Step 2 (Connection Analysis).}
    \label{fig:error_analysis_part2}
\end{figure}

\textbf{Human Intervention:} Our analysis of the interventions revealed two primary sources of error in the automatic conversion: (1)~perspective shifts: In 3 of 34 manuals, when a manual changed the drawing perspective between steps, the model struggled to perceive those angles and lost track of the same components across steps( Figure~\ref{fig:error_analysis_part2}A). In human intervention, this required indicating where the part has gone across steps in the manual. And (2)~spatial hallucination: When \name{} could not identify a clear visual match for an extracted part in the assembly diagrams, it selected a blank area as the predicted component location (Figure~\ref{fig:error_analysis_part2}B). This occurred in 2 of the 34 manuals. Human intervention required replacing these selections with the correct component locations. One manual contained both error types, so four manuals required correction in total.

\subsubsection{Step 3: Final Instruction Generation Achieved 91.2\% Success before Text Correction}
The final stage generates a structural roadmap and procedural instructions from the part relations produced in Step 2. 

\textbf{Success Criteria:} We reviewed the generated descriptions against the source visual
manual for unsupported statements, incorrect explanations, and omitted assembly details. An output met our technical success criteria when the review identified none of these errors.

\begin{figure}
    \centering
    \includegraphics[alt={Two examples show errors in generated instructions and their correction through text editing. In A, a manual diagram is paired with generated text that omits fastening locations; the reviewer adds that the locking nuts connecting the seat base to the leg frames should be tightened at the top and underside. In B, generated text includes a warning against using a power drill that is unsupported by the manual; the reviewer deletes it. Interface screenshots show controls for editing and saving the text.},width=1\linewidth]{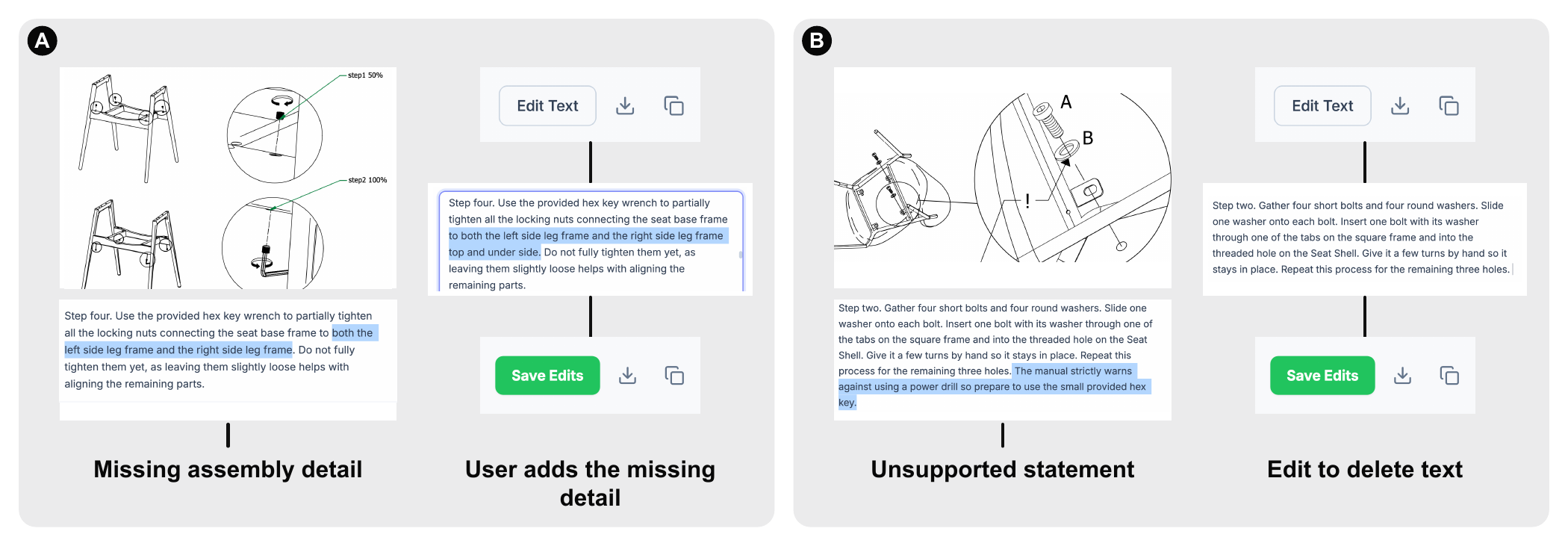}
\Description{Two examples show errors in generated instructions and their correction through text editing. In A, a manual diagram is paired with generated text that omits fastening locations; the reviewer adds that the locking nuts connecting the seat base to the leg frames should be tightened at the top and underside. In B, generated text includes a warning against using a power drill that is unsupported by the manual; the reviewer deletes it. Interface screenshots show controls for editing and saving the text.}
    \caption{Failure modes during the automated processing stages of \name{}. Cases A and B show missing assembly detail and unsupported statement during Final Instruction Generation with suggested editing steps}
    \label{fig:error_analysis_part3}
\end{figure}

\textbf{Evaluation Results:} Given the corrected outputs from the preceding stages, instructions for 31/34 manuals (91.2\%) met our technical success criteria. The remaining three outputs required correction. Two contained statements unsupported by their source manuals. One asserted that the manual prohibited using a power drill and required the provided hex key, although the source manual contained no such warning. Another instructed users to obtain another person's help to turn the desk upright, although the source manual did not specify that assistance was needed. The third output omitted a detail about tightening the locking nuts from both the upper and lower sides of the assembly. Figure~\ref{fig:error_analysis_part3} illustrates an unsupported statement and an omitted assembly detail.

% to finish up with some of the failed cases
\textbf{Human Intervention:} The interface allows reviewers to delete unsupported statements and add or revise assembly details before saving the corrected instructions. The illustrated corrections involve removing an unsupported warning and adding missing fastening information. Opening the editor and saving the revised text are interface actions. Across the three outputs, which contained 1,662 words before correction, reviewers deleted 25 words and added 4 words.

Across 34 manuals, \name{} met our technical success criteria before correction at the respective stage for 32/34 manuals (94.1\%) in part extraction, 30/34 (88.2\%) in connection analysis, and 31/34 (91.2\%) in instruction generation. We recorded failures separately for each stage but did not track which manuals required correction across multiple stages. After human review and correction, outputs for all 34 manuals (100\%) met the technical success criteria. These results assessed correctness relative to the source manuals. To evaluate the quality of the instructions we conduct qualitative analysis with BLV individuals in our user study.

\section{User Study: Roadmap Use and Physical Assembly Strategies}
We studied how BLV individuals work with the instructions \name{} generates, and what strategies they bring to the physical assembly. Since no comparable accessible instructions exist for flat-packed furniture, the study characterizes this experience rather than benchmarking it against a baseline. We measured workload and followed up with semi-structured qualitative interviews. Sections~\ref{sec:study-participants} through~\ref{sec:study-analysis} describe the method; Section~\ref{sec:study-findings} reports our findings.

\subsection{Participants}
\label{sec:study-participants}
Inclusion criteria for our study: participants were required to (1) self-identify as blind or low vision, (2) be at least 18 years old, (3) be proficient in using screen readers, and (4) have sufficient upper-body mobility to manipulate objects with both hands. We recruited eight participants (six female, two male) through organizations serving blind people, the research team's professional networks, and snowball sampling. 
Table~\ref{tab:participants} contains a summary of the participants' demographics. Participants were compensated 80 USD for 1.5 - 2 hours of study. The recruitment and
study procedures were approved by the Institutional Review Board
(IRB anonymized for review). %protocol number IRB0150454.

\begin{table*}[b]
\AccessibleTableDescription{Demographics and assembly experience of eight participants. Columns give participant identifier, age, gender, vision level, acquired or congenital vision loss, and assembly experience during the previous two years. P1 and P8 are totally blind; P2 through P7 are legally blind with light perception. P1: male, 62, acquired vision loss, one unsuccessful assembly attempt. P2: female, 57, acquired, one unsuccessful attempt without instructions. P3: female, 23, congenital, a few attempts with difficulty handling small parts. P4: female, 55, congenital, no assembly experience. P5: male, 24, congenital, one attempt with assistance identifying parts. P6: female, 36, acquired, no experience and reliance on family. P7: female, age 25 to 34, acquired, one attempt with guidance. P8: female, 20, congenital, no experience. P7 disclosed an age range rather than an exact age.}
\centering
\small
\begin{tabular}{p{0.2cm} p{0.8cm} p{0.6cm} p{4cm} p{1.2cm} p{6cm}}
\toprule
\textbf{ID} & \textbf{Age} & \textbf{Gender} & \textbf{Vision Level} & \textbf{Vision status} & \textbf{Assembly Experience in past two years} \\ \midrule
P1 & 62 & M & Totally Blind & Acquired &  1 unsuccessful attempt \\
P2 & 57 & F & Legally Blind with light perception & Acquired & 1 unsuccessful attempt; without using instruction \\
P3 & 23 & F & Legally Blind with light perception & Congenital & A few attempts; Struggled with small parts  \\
P4 & 55 & F & Legally Blind with light perception & Congenital & None  \\
P5 & 24 & M & Legally Blind with light perception & Congenital & 1 attempt with assistance for part identification \\
P6 & 36 & F & Legally Blind with light perception & Acquired & None; Relies on family  \\
P7 & 25-34* & F & Legally Blind with light perception & Acquired & 1 attempt with assistance for guidance \\
P8 & 20 & F & Totally Blind & Congenital & None; \\
\bottomrule
\end{tabular}
\caption{Demographics and relevant experience of participants. *P7 disclosed an age range instead of a specific age.}
\label{tab:participants}
\end{table*}

\subsection{Apparatus}
\begin{figure}[t]
\centering
\includegraphics[alt={Four photographs compare furniture parts before assembly with assembled furniture. The top row shows an office chair, including its seat, backrest, five-legged base, casters, and gas lift. The bottom row shows a four-tier shoe rack made from metal tubes, plastic connectors, and fabric shelves. Parts are arranged on a table in the left column; a participant handles the assembled furniture in the right column.},width=0.75\linewidth]{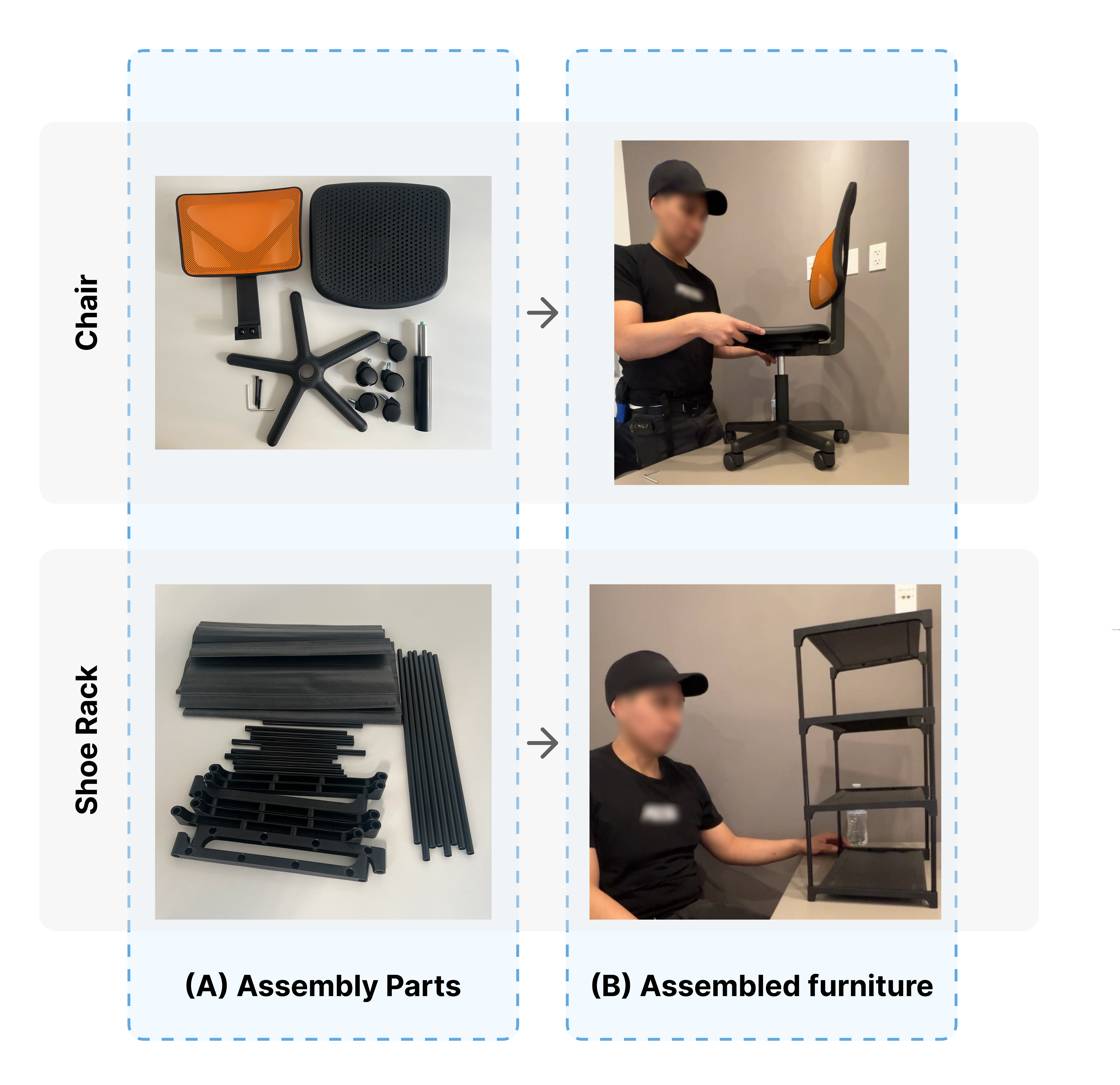}
\Description{Four photographs compare furniture parts before assembly with assembled furniture. The top row shows an office chair, including its seat, backrest, five-legged base, casters, and gas lift. The bottom row shows a four-tier shoe rack made from metal tubes, plastic connectors, and fabric shelves. Parts are arranged on a table in the left column; a participant handles the assembled furniture in the right column.}
\caption{The two pieces of furniture participants assembled. (a)~the parts as we laid them out for participants and (b)~the furniture assembled by one participant.}
\label{fig:furniture}
\end{figure}

Participants assembled two pieces of flat-packed furniture (Figure~\ref{fig:furniture}). To control for sequence effects, the tasks were counter-balanced across the eight participants. We chose a desk chair and a shoe rack model because they are common household items. They have different assembly principles, and both models are available on Amazon with positive reviews and are affordable (< 45USD). We attached the full instruction protocol in our appendix~\ref{appendix:instructions}.
\begin{enumerate}
    \item \textbf{Shoe Rack:} A four-tier shelving unit requiring the repetitive insertion of long and short metal tubes into four-way plastic connectors, wrapped in tensioned fabric sheets. This task requires repetitive procedural logic, topological orientation, and the management of multiple identical parts. It required no tools.
    \item \textbf{Office Chair:} An office chair requiring the correct orientation of a five-star base, the friction-fit insertion of five casters (wheels), the placement of a gas lift, and the use of a hex key (Allen wrench) to bolt a heavy backrest to a seat cushion. This task requires spatial manipulation, bi-manual coordination, tool usage, and asymmetric part alignment.
\end{enumerate}

\subsection{Task}
We provided participants with the parts of each piece of furniture laid out as in Figure~\ref{fig:furniture}a; they received instructions generated by \name{} which they accessed through their preferred screen reader. We encouraged participants to think aloud; they were free to revisit each instruction at their own convenience. We had one researcher in the room, who intervened to prevent physical damage to the components (participants did not hurt themselves) or to provide requested assistance if participants did not know how to proceed. We distinguished minor assistance with locating or retrieving parts from critical assistance with progressing through or physically performing an assembly step, as defined below. Following each assembly, participants completed the NASA Task Load Survey (NASA-TLX) to quantify subjective workload across several dimensions, followed by a semi-structured interview to capture deeper qualitative reflections. The full interview guide is available in the Appendix~\ref{appendix:questions}.

\subsection{Metrics}

We assessed subjective workload using a seven-point adaptation of the NASA-TLX. After each assembly task, participants rated mental demand, physical demand, temporal demand, effort, frustration, and perceived performance on scales from 1 to 7. We report each dimension separately to provide descriptive context for participants' assembly experiences.

We recorded minor assistance and critical assistance separately. Minor assistance included participant-requested help with retrieving dropped parts or locating parts in the workspace. Critical assistance was recorded when participants were unable to resolve an assembly step for more than 10 minutes or required physical help with assembly.

Finally, we counted the mistakes that the participants made during the assembly. Mistakes included forgetting to assemble parts, assembling parts in the wrong direction, and loose connections during alignment. We count both during the assembly, when participants spend a long time without noticing the error, and at the end of the study, when participants may still be unaware. Table~\ref{tab:mistakes} identifies participants who made each type of error during assembly, including errors they subsequently corrected. 

\subsection{Data Collection and Analysis}
\label{sec:study-analysis}
Our data included audio and video recordings, text transcripts of study sessions, and NASA-TLX scores. A thematic analysis~\cite{braun2006using} was conducted. One researcher first transcribed the video, then reviewed the full interview transcripts and performed open coding. Initially, all sessions were coded, and the resulting codes were then organized through two rounds of affinity diagramming to develop the themes and categories presented in this paper. A total of 15 categories were initially identified. Through discussion among the authors, we developed them into three themes that directly address our research questions regarding non-visual instruction design (RQ1) and BLV assembly strategies (RQ2).

\subsection{Findings}
\label{sec:study-findings}
We first report what the participants did: whether they completed the assemblies, where they needed help, the errors they made, and how they rated the workload. We then report the three themes our thematic analysis produced, developed from behavioral observations and think-aloud transcripts and contextualized by the post-hoc semi-structured interviews.

\subsubsection{Assembly Outcomes and Interventions}

\begin{figure}[h]
\centering
\includegraphics[alt={Sixteen workflow diagrams show eight participants' first and second assembly tasks, one shoe rack and one chair each. Horizontal position represents progress through instructions read aloud by a screen reader; vertical position represents physical assembly progress. Purple paths represent the rack and cyan paths the chair. Numbered events trace listening ahead, revisiting instructions, performing assembly steps, and undoing or repeating steps. Exclamation marks indicate critical assistance, lightning symbols indicate minor assistance, and circular arrows mark repeated attempts. Critical assistance appears in P3's chair task and both P7 tasks. Rack sessions lasted 13 to 37 minutes and chair sessions 8 to 36 minutes.},width=0.8\linewidth]{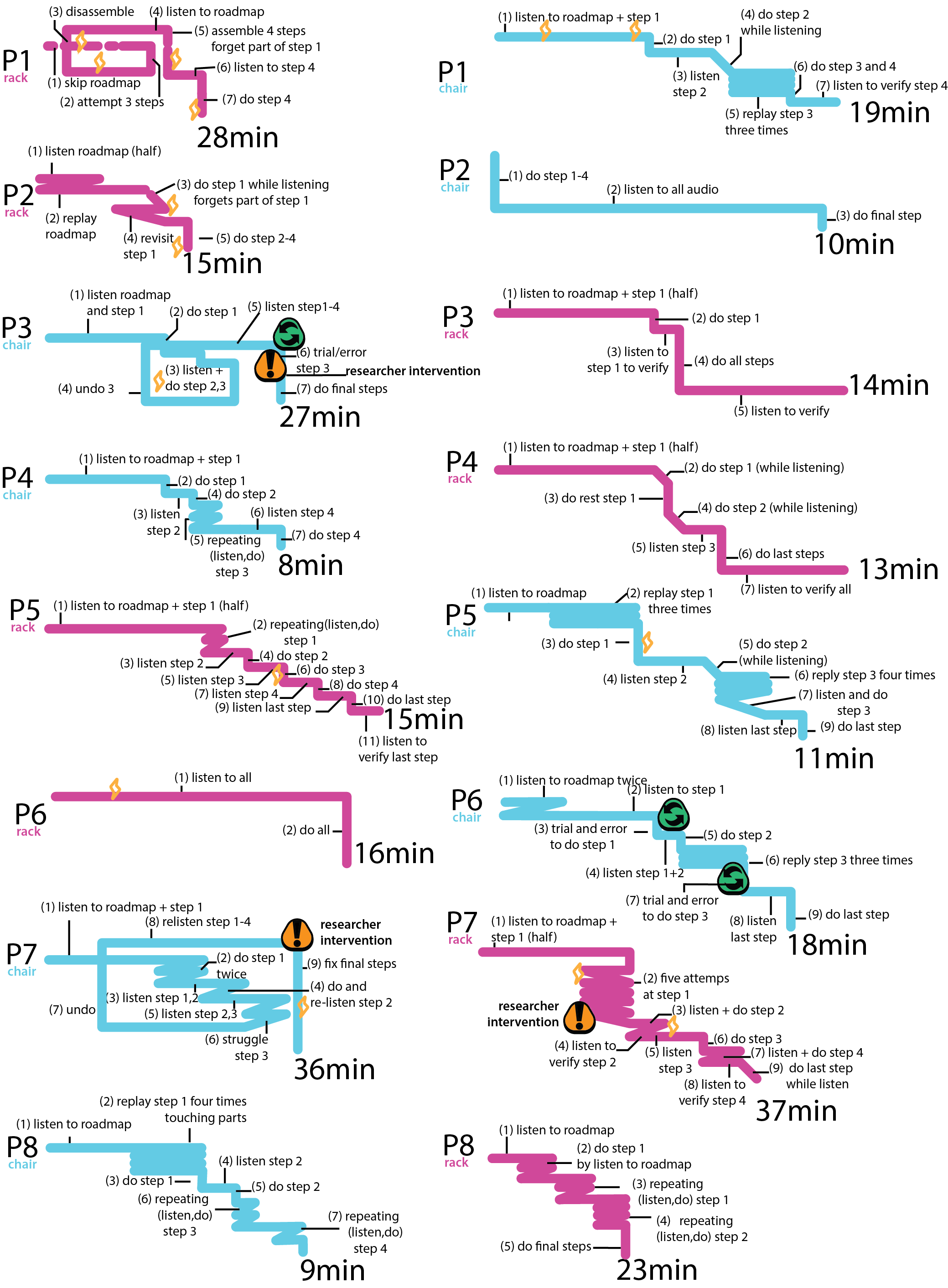}
\Description{Sixteen workflow diagrams show eight participants' first and second assembly tasks, one shoe rack and one chair each. Horizontal position represents progress through instructions read aloud by a screen reader; vertical position represents physical assembly progress. Purple paths represent the rack and cyan paths the chair. Numbered events trace listening ahead, revisiting instructions, performing assembly steps, and undoing or repeating steps. Exclamation marks indicate critical assistance, lightning symbols indicate minor assistance, and circular arrows mark repeated attempts. Critical assistance appears in P3's chair task and both P7 tasks. Rack sessions lasted 13 to 37 minutes and chair sessions 8 to 36 minutes.}
\caption{The workflows of participants. The horizontal axis depicts progress through instructions read aloud by a screen reader and the vertical axis progress in physical assembly. Purple graphs are the rack and cyan graphs the chair assembly. exclamation marks indicate critical assistance with assembly and thunder icons indicate minor assistance with locating or retrieving parts. The recycle icon is where a user repeatedly tried to assemble the same part.}
\label{fig:workflows}
\end{figure}

All eight participants completed both assembly tasks. The level of autonomy varied. Figure~\ref{fig:workflows} shows an overview of how participants used the instructions during the assembly. Four participants required one or two minor interventions (depicted in the workflows by thunder icons), primarily for locating parts on the workspace or retrieving parts they dropped on floors. 

Critical assistance occurred in three assembly sessions involving two participants (depicted with an exclamation mark): P3 struggled with physical strength for aligning the seat and the backrest. P7 required logical guidance after being unable to progress for 10 minutes on both the chair and the shoe rack assembly tasks. We see some sequence effects where participants skipped the roadmap instructions on their first assembly and then paid more attention during the second assembly resulting in a smoother workflow~(e.g., P1,P3).  

% \begin{figure}[p]
% \centering
% \includegraphics[
%     width=\textwidth,
%     height=0.90\textheight,
%     keepaspectratio
% ]{figures/moneyshot911.png}
% \caption{Participants' assembly workflows. Each row shows one
% participant's first and second assembly sessions. Colors,
% symbols, and axes are explained in the figure legend.}
% \label{fig:workflows}
% \end{figure}

\paragraph{\textbf{Analysis of Assembly Errors}}
We observed assembly errors and misalignments across both tasks (Table~\ref{tab:mistakes}). Some errors were subsequently corrected, while others remained at task completion. We grouped the observed errors into three categories:
\begin{itemize}
\item \textbf{Structural Errors} Most mistakes in the rack assembly are flipping of the fabric sheet and forgotten layers. 
\item \textbf{Fasteners} Issues with loose connecting tubes in the rack and bolt connections in the chair task were common. These steps often require participants to use their bodies to stabilize the structure during alignment. 
\item \textbf{Orientation} A common orientation error happened when participants attempted to turn the caster into their base instead of press-fitting them into the assembly.
\end{itemize}

\begin{table}[t]
\AccessibleTableDescription{Observed assembly errors are grouped into structural errors, fastener errors, and orientation errors, with affected participants listed for each. Structural errors: omitted fabric sheet, P1 and P2; upside-down fabric sheet, P4, P5, P6, and P7; upside-down shoe-rack layer, P7. Fastener errors: inserted bolts before connecting the seat and backrest, P3, P6, and P7; loose bolt or tube connections, P2, P3, P5, P7, and P8. Orientation error: turned a caster instead of pressing it into place, P1, P4, P6, and P7. Participants may appear in multiple rows. The table includes errors that participants subsequently corrected.}
\centering

% Standard tabularx permits semantic Table/TR/TH/TD tagging on arXiv TL2025.
\begingroup
\renewcommand{\tabularxcolumn}[1]{m{#1}}
\renewcommand{\arraystretch}{1.364}
\setlength{\tabcolsep}{6pt}
\begin{tabularx}{\linewidth}{|>{\hsize=.5\hsize\linewidth=\hsize\raggedright\arraybackslash}X|>{\hsize=1.5\hsize\linewidth=\hsize\raggedright\arraybackslash}X|c|}
\hline
\rowcolor{gray!20}\textbf{Category} & \textbf{Description} & \textbf{Participants} \\ \hline
\ACMTableRowSpan{3}Structural\newline Errors & Forgot to assemble fabric sheet & P1,P2 \\ \hline
& Assembled the fabric sheet upside down & P4,P5,P6,P7 \\ \hline
& Assembled the shoe rack layer upside down & P7 \\ \hline
\ACMTableRowSpan{2}Fasteners & Assembled bolts before connecting the seat and backrest & P3,P6,P7 \\ \hline
& Loose connection in bolts or tubes & P2,P3,P5,P7,P8 \\ \hline
Orientation & Turned caster instead of pressing & P1,P4,P6,P7 \\ \hline
\end{tabularx}
\endgroup
\caption{Assembly errors observed during the study, including errors subsequently corrected. Participant IDs indicate for whom each error type was observed.}
\label{tab:mistakes}
\end{table}

\paragraph{\textbf{“IKEA-Effect”}}
Participants, including the ones with faulty assembly, rated their performance score between 4 and 7, with six participants rating themselves above 6. This may be caused by the so-called “IKEA-Effect”~\cite{norton2012ikea} in which consumers disproportionately value products they assembled, despite imperfections. This foregrounds the value of autonomous assembly for BLV individuals.

\subsubsection{Workload}

\begin{figure}[h]
    \centering
    \includegraphics[alt={Twelve horizontal stacked bars compare shoe-rack and chair ratings for six dimensions of the adapted NASA Task Load Index: mental demand, physical demand, temporal demand, effort, frustration, and perceived performance. Each bar represents eight participants; numbers inside segments indicate participant counts. The color legend runs from blue for positive responses to red for negative responses. Performance ratings cluster toward the positive end, while mental demand and effort show more mixed responses.},width=0.4\linewidth]{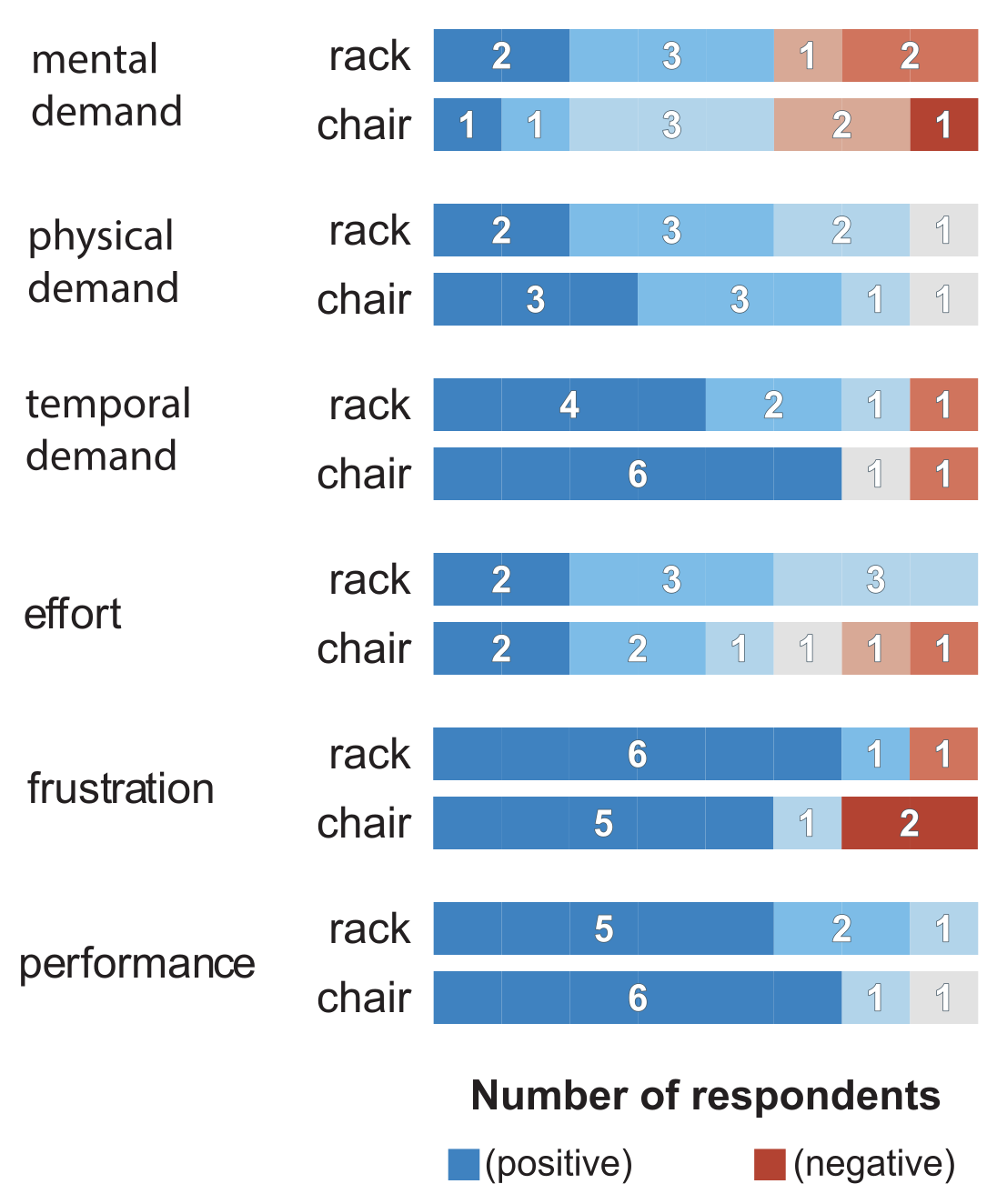}
\Description{Twelve horizontal stacked bars compare shoe-rack and chair ratings for six dimensions of the adapted NASA Task Load Index: mental demand, physical demand, temporal demand, effort, frustration, and perceived performance. Each bar represents eight participants; numbers inside segments indicate participant counts. The color legend runs from blue for positive responses to red for negative responses. Performance ratings cluster toward the positive end, while mental demand and effort show more mixed responses.}
    \caption{Distributions of ratings for the six dimensions of the adapted NASA-TLX after each assembly task (\AccessibleInlineMath{n equals eight}{n=8} per task). Numbers within the segments indicate how many participants selected each rating.}
    \label{fig:scales}
\end{figure}

On average, participants spent 20.1 minutes assembling the shoe rack and 17.25 minutes on the chair, while the participants' perceived effort on the chair~(3/7) slightly exceeded that of the shoe rack~(2.125/7). Frustration was higher for the chair assembly~(2.75/7) than for the shoe rack~(1.75/7). All results of the NASA-TLX are reported in Figure~\ref{fig:scales}. With eight participants we report these scores descriptively, to contextualize the qualitative findings that follow rather than to support statistical inference.

Workload ratings did not always correspond to observed assembly errors. P1 for example, made several mistakes on the shoe rack assembly, such as incorrectly building a frame with all long tubes, and forgot to assemble the fabric sheet, but rated the perceived workload~(mental, physical, temporal, efforts, frustration) very low while rating their performance high. They were aware of the mistakes but stated these were caused by the task being too easy “It was relatively easy to me, so that's why I, overlooked the shelf.” 
% Most participants state their temporal and frustration with the minimum score as they finally can complete the tasks. This could suggest the mistakes they faced during the process are more like trial and error which challenges their thinking but not dramatically affect their feelings. 

In contrast, P6 corrected all errors made during assembly and completed both furniture items correctly. Nevertheless, P6 rated temporal demand as 6 out of 7 for both tasks, and frustration as 6 out of 7 for the shoe rack and 7 out of 7 for the chair, suggesting that achieving a correct final assembly could still involve substantial perceived workload.

\subsubsection{Theme 1: Spatial Cognition and Mental Mapping}
We observed that participants who carefully paid attention to the roadmap instructions prior to assembly (P3, P4, P5, P8) performed noticeably better with less hesitation and confusion during subsequent procedural instructions. Participants who initially skipped part of the roadmap (P1 initially skipped most of the roadmap during the shoe rack tasks, P3 initially skipped most of the roadmap during the chair task) had a hard time grasping the relation between instructions and struggled during procedural assembly leading to mistakes.

\paragraph{Roadmap Perspective Guided The Procedural Assembly}
We found a sequencing effect for P1 and P3, during their first task they were unaware that the roadmap would provide valuable information during the procedural assembly. They therefore paid more attention to this in their second assembly task. This reveals that users may adopt the instructions more fully once they realize the value. 

P3 contextualized this in their reflection after assembling the shoe rack (second task) \textit{“The instructions give me description of how it is supposed to look and which piece does what. Looking back I wished I did this more with the chair because I have a really good idea this piece is this, this piece is this. Naturally I just try to figure it out myself, I'm not a huge instruction reader, but what is great that it has instructions on all the parts and how it is supposed to look at the end, that helps a lot. Because now even without reading the details of instructions, I know these parts~(small tube) go here~(pointing to corner of the layer). But I know that mainly because at the beginning told me the overview.” }
%P1 also share similar thoughts on he could get rid of instructions after understanding the structure, “I do the instructions first, and especially those tell me the parts. and if i started to do step by step thing but then as it go along, i don’t need to go to step by step things(instructions) unless i ran into the problem. why it is not holdings, why it is not tight, why it is not fitting.”
Participants frequently referred to the roadmap as a “recipe” (P5), “map” (P3, P5), “picture” (P3, P4, P5, P6). P5 stated it helped them to later “piece together efficiently and then at a good pace to without getting confused”, P4 valued that it makes them  “Think what things should look like”. 

\paragraph{Errors from Missing the Roadmap Perspective}
When P1 assembled the shoe rack, they listened to a few words of the roadmap, realized these were not procedural instructions, so they skipped these and jumped directly into the detailed procedural instructions. In those instructions, they identified key words “shelf” and “long tubes”, and forgot they should build a shoe rack. They inferred from those words this is a “long shelf” so they stacked all long tubes with connectors together. When they ran out of tubes, they realized this is not what they were supposed to build. They then returned to the beginning of the instructions and realized this should be a shoe rack. We observed similar tendencies in P2 and P3, who sometimes proceeded with assembly before hearing all of the procedural guidance, apparently relying on their existing understanding of the task.

The sequence effect found in our study suggests that after participants realized the utility of the roadmap from their first assembly, they felt well supported on their second assembly, by carefully listening to the roadmap descriptions.

\paragraph{Visual/Haptic (Mis)alignment}
Discrepancies between a participant's structural understanding and their tactile experience can lead to skepticism regarding the accuracy of subsequent instructions. We identified several 'implicit defaults' within the product design that are not explicitly addressed in the manual, creating further uncertainty for the user. \name{} currently either uses part names from the manuals if provided or generates a part name based on their shapes. For example, in the desk chair visual instructions, its rotatable wheels were called \textit{caster}, while this is directly related to its shape and function, P1, P3 ,P6, and P7 raised questions about which parts the casters were. 

Translating instructions without having access to the physical assembly may omit information crucial to users. We observed that fabric sheets for the shoe rack have two sides, one side is smooth, which is considered to be the front side. Since this isn’t mentioned in the visual instructions, \name{} also missed out on this. However, some participants noticed that the sheets have two sides. P3, P4, P5, and P6 suggest using tactile descriptions for orientation, P5: “the part that feels more like rubber should be facing up or describe the textures of the hidden flaps” and P6 suggested: “plastic parts should have been on top, while the stitching line should be hidden facing the floor”. 

In addition to fabric sheet, the shoe rack connectors have four side holes. During assembly, only two side holes are used. Visual instruction doesn't mention this feature in diagrams or text which made \name{} unable to account for this. P4 and P6 noticed the plastic connectors for the shoe rack had two unused holes in the middle of each connector. Both participants spent considerable time in understanding these four holes' functions and reasoning where they should attach their two long tubes. They later realized they should attach long tubes to outer holes, P4 stated “I don't understand about those holes... nothing there, they don't fit in there anyway”. P6 concluded these two unused holes as “not an instruction problem. That's a product problem”. We assume these holes are there to simplify manufacturing and make parts interchangeable. It would be useful for tools like \name{} to warn users about such problems. Mentioning the features that go unused may be just as important as mentioning the used ones to avoid confusion or distress of missing parts.

\subsubsection{Theme 2: Embodied Ergonomics and Kinetic Semantics}
A large theme we found in our analysis pointed to the role of embodied ergonomics as an additional modality for BLV assembly, we observed participants intuitively gravitate towards that, pointing to exciting potential for future assembly support.

\paragraph{Using Proprioception and Haptic Feedback to Align}
In the chair task, \name{} translated the visual instructions to “Lay the seat upside down on a flat surface so the bottom faces up. Bring the lower extension of the back against the rear edge of the seat so their connection holes line up.” All participants struggled with this alignment, P3 tried to follow the instructions but couldn’t align them while holding the bolt. The observer intervened by providing help to hold the backrest and the seat. P5 tried to assemble them on the table but they experienced significant friction when connecting the backrest to the seat. They later realized they needed to “slightly lift it up” to align the connectors and bolts. P4 suggested instructions should proactively tell the user when they need to lift or angle parts in the air to align bolts, rather than assuming they will naturally figure it out while the parts are flat on a table.  

We observed P1, P5, P6, P8 use their legs to replace the flat surface to align parts, as mentioned by P6 “If I'm punching both parts... I have sensory feedback from both parts. So because of that... if it moves on my lap, I'm gonna know faster than if it moves on the table because it's touching my legs”. Participants used their bodies as a supportive structure to obtain continuous tactile feedback. Additionally, they can fasten the seat on their legs by squeezing it with their legs. P1 tried to put the seat vertically between their legs to stabilize it. P4 applied a similar strategy because they “couldn't get good grips” on the table. P4 later reflects “I didn't feel comfortable [on the table], I feel more comfortable holding it... I just use my body as pose help. It helped me to hold parts together, I use my body as a lever”. Our observations suggest that BLV instructions could exploit this embodied scaffolding—the use of the lap, chest, or legs as a supportive structure which provides tactile feedback while leveraging the relative proprioceptive feedback of arms. 
% Action like alignment should be considered as "process" rather than a "final state"; instructions should describe the tactile "pre-conditions" and the expected haptic confirmation after alignment complete. 

\paragraph{Kinetic Choice of Words}
We observed a tension between the verbalized mechanical actions in the instructions and the perceived affordances of the hardware. While certain participants (P3, P5) strictly adhered to the audio cues, the majority (6/8) prioritized tactile exploration, which often led to "tactile-first" errors. For instance, when instructed to “Press one caster firmly into the hole,” several participants (P1, P4, P6, P7) instinctively attempted to screw the part in. The caster's affordances guide users that way, the shape includes a stem that fits into the hole in the base which leads to an assumption on how things are supposed to be fastened. This latent understanding overrides our instructions. P4 and P6 only corrected their actions after attempting to brute-force rotate the caster into the base, P6 states "didn't pay attention" to "press firmly".

% After this failed, participants touched the hole and the caster to confirm whether/how they fit. P4 and P6 tried rotating the caster in two directions, and then brute forced them into the base. They later realized they should simply “press” them into the base. P4 reflected they need to pay more attention to the instructions “If I pay better attention I would know wheels are not roll in”.  P1 and P2 after having difficulties, revisited the instructions to get a sense of what they missed. During the semi-structured interview, P6 said they didn’t pay attention to “press firmly” and almost give up on this step.

We observed when participants played back the instructions and found useful information, they nodded their heads or repeated the words confirming they understand the actions. Specifically, we see relational words like (“through”, “side holes”) and comparative words like (“wider part”), and force-related action phrases such as ‘press firmly’ serving as high-fidelity cues that allowed participants to resolve “stage-specific” bottlenecks. They either provide a direction or comparison with other parts, helping them confirm subsequent actions. P3 also suggested that instructions should explicitly state the purpose of hardware as well rather than just the action. 

\paragraph{Tool and Small Parts Friction}
During the backrest and seat alignment task, we observed 5/8 (P1, P3, P4, P6, P7) participants having difficulty using an Allen key to tighten the bolts. P1, P4, P6 and P7 dropped the tool multiple times, P1 later mentioned: “sometimes when you turn it, it comes down. So you have to, you know, and get it in there. And so just keep it. But sometimes it, it may be a problem”. Similar to our observations around the alignment of parts on participants' legs, we observe that the alignment of a tool with a bolt is challenging. 

We asked participants why they don’t assemble furniture on their own. The main point was that instructions are highly visual so they don't know where parts need to go (P1, P3-P8), but participants also expressed fear of “too many small or delicate parts” getting lost. P3 felt relief when assembling the shoe rack because they “don't even have to deal with the whole screws”, P3 added “even with perfect instructions, the physical affordances of the hardware itself (e.g., friction-fit tubes vs. tiny screws) drastically impact the accessibility of the task.”

% \paragraph{unnoticeable information from instructions}
% Translating instructions without having the physical assembly may miss out on crucial information to users. We observed that fabric sheets have two sides, one side is flat, which is considered to be the front side. Since this isn’t mentioned in the visual instructions, our accessible instructions also missed out on this fact. However, some participants noticed that the sheets have two sides. P3,P4,P5, and P6 suggest using tactile descriptions for orientation, P5: “the part that feels more like rubber should be facing up or describe the textures of the hidden flaps” and P6 suggested: “plastic part “should have been on top, while the “stitching line” should be hidden facing the floor”. 

% We therefore consider \name{} to be primarily used by the designer/manufacturer who distributes the instructions rather than end-users who may miss out on these details. The authoring functions allow designers to include this in their instructions. 

\subsubsection{Theme 3: Assembly Workflow and User Agency}
We observed how participants successfully managed the workspace and parts, so they can be independent during the assembly. Participants made suggestions on how to improve our instruction to reinforce their independence. 

\paragraph{Workspace and Inventory Management}
Visual instructions assume users can visually check any leftover parts on the table so people can track current progress. However, as previously stated for P1 and P2, they only knew they forgot to install fabric sheets for shoe racks when they believed they finished the assembly. We observed some participants (P2,P4) sweep the entire workspace to make sure no parts were left unassembled. P7 and P8 confirmed their final assembly by feeling the chair “it started to move like a chair.”

One suggestion from participants is to provide the part list in the beginning of each assembly step (P3) or manual page (P6, P7). \name{}'s instruction provides a part list only at the beginning of the instruction. Providing it closer to current actions can reduce the burden of scrolling back and forth. As suggested by P7: “Visually, I think people just always have the picture as you do each thing and usually gives you another diagram. So maybe keep this on every page, so you don't have to scroll all the way up”. P3 also wanted to know the total number of items at the beginning of a step so they can count their inventory before starting the next part.  

\paragraph{Instruction Design Considerations}
Participants often requested specific descriptions to identify part orientation and functions. P1 suggested describing a plastic connector as a "rectangle" with "four holes" to help users identify it through a process of elimination and even "overkill" for part description but create a clear image for part's function. Similarly, when P5 were uncertain about the orientation of a connector, they asked if the instructions could verify that the connector should look like an "n" rather than a "U". This aligns with findings in “Lost in Instructions”~\cite{reddy2026lost}: BLV participants wanted any kind of orientation cues to be helpful while in the assembly process.

The audio instructions supported BLV individuals, especially P2 compared their previous experience when only having printed visual manuals they couldn't read. P5 praised the digital format of the instructions, explicitly noting that they found the process "seamless" because they could easily "scroll through them with the keyboard" rather than struggling to navigate a custom mobile app. P3 recommended changing the current paragraphs to bullet points to aid screen-reader navigation. The built-in Mac VoiceOver screen reader requires the user to press complex multi-key combinations (e.g., Control + Down Arrow), which occupies both hands. Bullet points act as smaller, easily scannable tasks that allow skipping through the document using one hand while holding parts for assembly with the other. She stated “it is more scannable to me… I have to stop for the step by step, if it was bullet point, it has already broken down.” P7 also suggestted breaking each step down to be one action only so they can better remember all actions in that step. 

\paragraph{Independence Support}
We observed participants wished to be independent during the assembly. P3 and P5 described feeling accomplished after completing the assembly tasks. P2 only consulted the audio when they got stuck on assembling casters because this is “something that I am doing for myself”. P7 also brought their smart glasses to the study and wished to use it because it can tell if some parts of assembly are tilted “Because it would have told me that I didn't have it even (the shoe rack) instead of me having to figure it out myself. Visually, it would have been like you're in the wrong hole. Because that's a lot of holes to have to fill with your two hands.”

\paragraph{From Avoidance to Therapeutic Enjoyment}
Making the assembly process accessible led to a shift in affective response. For many participants, furniture assembly was associated with a "fear of failure" or assumptions that it would be inaccessible. In contrast, during the study, P8 described the process as it “felt like really the therapeutic, like you were doing some art or craft project because all the instructions made sense. No, I was not annoyed or frustrated about it.”

\section{Discussion}
Our findings suggest that making assembly instruction accessible requires more than a direct translation of visuals. It is essential to reconstruct information to improve non-visual reasoning. In this section, we reflect on our two core RQs. To address RQ1, we discuss the necessary development in manual designs and design principles for translating visual instruction to survey knowledge. To address RQ2, we reflect on the role the roadmap played, and on the physical assembly strategies BLV individuals developed, including embodied ergonomics, error recovery, and verification. We conclude by converting these discussions into concrete guidelines for assembly instructions.

\subsection{Translating Visual Instructions to Accessible Instructions}
In exploring how to develop effective instructions (RQ1), our development of the tech probe highlighted limitations in directly passing visual manuals to BLV participants. Supporting the development of survey knowledge depends on how information from visual manuals is selected, organized, and translated into accessible descriptions. Here we lay out insights regarding this process.

\subsubsection{Expanding The Design Process}
We identified several opportunities to improve the initial design of assembly manuals to better support both automated translation and users' understanding.
\begin{enumerate}
\item \textbf{Integrate text during the initial design phase.} Our current probe translates visual instructions only after the product and manual have already been finalized. In our benchmark of 34 manuals, only two included any textual instructions. Given that most manuals leave significant white space around assembly diagrams, accessible, non-visual text descriptions could be authored natively during the manual's initial design phase rather than relying on post-hoc translation.

\item \textbf{Standardize visual cues.} During our phase-by-phase evaluation, we observed that multi-perspective drawings and severe angles confused our own researchers and, unsurprisingly, induced spatial hallucinations in the VLM. To explain such complex assemblies, designers should explicitly illustrate how parts connect using clear, dashed lines, alongside isolated reference images of the parts~\cite{seehorn2025beyond}. Standardizing these explicit visual cues reduces the cognitive load for users while significantly lowering the computational reasoning burden for VLMs.
\end{enumerate}

\subsubsection{Why The Translation Cannot Be Left To BLV Individuals Alone}
It might seem feasible to simply provide BLV individuals with a VLM to independently translate printed manuals. However, our exploration revealed logistical and safety hurdles that make direct translation impractical:

\begin{enumerate}
\item \textbf{Inconsistent manual design and image capture.} The translation process relies heavily on a good layout, such as clear part lists with page numbering, and distinct step indicators. However, our review of 34 diverse assembly manuals revealed a severe lack of standardization. Layouts vary wildly, ranging from documents dominated by multilingual warning texts to dense diagrams compressed onto a single page. Furthermore, requiring a BLV user to capture well-framed photos of these complex documents introduces a significant accessibility barrier before the assembly even begins.
\item \textbf{Algorithmic hallucination and safety risks.} While VLMs can successfully interpret many visual elements, they remain prone to algorithmic hallucinations and incomplete semantic grounding, and to systematic biases that persist even on images the model appears to read correctly~\cite{vo2025vision, gonzalez2025towards, xie2025beyond}. If a VLM incorrectly infers a part connection and a BLV individual cannot visually verify the error, the resulting structural failure could pose a severe physical safety risk. BLV individuals moreover tend to attribute such failures to themselves rather than to the system~\cite{10.1145/3746059.3747597}, which makes an unverified pipeline doubly costly. Therefore, a mediated verification step is critical to ensure the integrity of the non-visual instructions.
\end{enumerate}

\subsubsection{Choice of Language}
Not all users share the same vocabulary when it comes to assembling furniture. In our formative study, we iteratively went from very technical language towards more general-purpose language. We also found that many words and descriptions of parts make reference to visual logic which is better avoided for BLV individuals~\cite{reddy2026lost}. While our goal was to serve a broad audience with \name{}, more fine-grained information about individual users could allow for fine-tuning to the specific vocabulary a given user is familiar with. 

An important aspect when providing structural roadmap as well as procedural knowledge is that terminology is applied consistently. In visual instruction, designers of manuals control the viewing perspective to give each part a consistent appearance to reduce confusion~\cite{agrawala2003designing, heiser2003cognitive}. The same applies to consistency between words and phrases used in procedural instruction and in structural roadmap.

\subsection{Designing for Survey and Procedural Knowledge}

Addressing the role of the structural roadmap (RQ2), we found that procedural instructions for BLV individuals are best followed when people have top-level survey knowledge of their task. It gives users the equivalent of browsing through a visual assembly guide before starting the actual assembly process. Our observations suggest that the roadmap helped some participants interpret subsequent instructions and avoid errors. That said, the roadmap requires users to hold information in mind while assembling, which itself imposes cognitive load. It is therefore crucial to be selective in what information to include and use language that is easy to parse without relying on visual reference.

Our study showed a variety of ways of consuming the instructions. Some skipped the roadmap at first and started to assemble right away, while others carefully listened upfront, and yet others used the roadmap at the very end to verify their assembly. Unlike visual instructions where a reader can glance back and forth to recall the overview, the sequential nature of audio makes it harder to navigate. Our observation is in line with \citet{li2026adcanvas}, who argue for easier navigation between different levels of knowledge. We argue that future BLV assembly guides could facilitate this breadth of strategies by allowing users to toggle between different modes of instruction. Another opportunity for roadmap design is to couple it with a physical task such as laying out the parts similar to assembly tools like \textit{RoadKill}~\cite{abdullah_roadkill_2021} or \textit{Daedalus in the Dark}~\cite{chang2021daedalus}. Several participants already used their hands during the roadmap to familiarize themselves with the parts.

\subsection{Physical Assembly Strategies}
Participants did far more with the instructions than read and follow them. Here we discuss three strategies we observed: how they recruited their own bodies into the task, how they recovered from mistakes, and how they verified the result.

\subsubsection{Bodily Dimensions and Tactile Reference}

When observing the physical assembly strategies BLV individuals use for procedural instructions (RQ2), we found that strategies extend far beyond simply connecting parts with their hands, echoing the range of non-visual strategies \citet{matsumoto2025assembling} report for assembling 3D-printed objects. To verify dimensions, ensure alignment, and evaluate the assembly state, BLV individuals frequently made use of their bodies as physical scaffolding. Specifically, we noticed participants actively changing their body positions to create stabilization. For example, some participants flattened their legs or adjusted their knee height to create a workbench, while others used their thighs and feet to squeeze objects together. This embodied strategy allowed them to release their hands from a holding position, freeing them to focus entirely on detailed bimanual alignments or complex fastening operations. Current visual instructions and the accessible instruction directly translated from them assume that assembly occurs on a flat surface like a table or on the floor. However, such a static workspace provides no sensory feedback. The body, by contrast, provides it continuously, telling users immediately when a part slips out of place or seats correctly.

We identify an opportunity for better non-visual instruction to explicitly tap into this potential and reference parts based on how they feel or even include guided strategies to align parts based on users' bodies rather than assuming assembly on a table/the ground (as we had done). Haptic systems built for blind users~\cite{schneider_dualpanto_2018} and the long line of haptic assembly simulation in manufacturing~\cite{xia_review_2013} offer starting points. Further research could explore how to map visual alignment strategies to haptic ones to better support this technique.

\subsubsection{Error Recovery}
When BLV participants made errors during assembly, recovering from those mistakes introduced significant friction. Based on our observations, this difficulty stems from two primary challenges: diagnostic backtracking and physical disassembly. Both are instances of the recovery and reversal problem long recognized in interactive systems~\cite{jr_user_1984}, but here the state to be reversed is physical.

First, errors were typically discovered only when the current physical state of the assembly became incompatible with a subsequent procedural instruction. However, diagnosing exactly when and where the error originated is uniquely challenging without visual reference. While sighted users can quickly check earlier diagrams to spot a mistake, BLV users cannot easily close such a verification loop through sequential audio. Participants therefore resorted to replaying the entire audio instruction or completely disassembling their work to reset their structural understanding.

Second, the physical mechanics of error recovery are often overlooked by product designers. Many furniture parts utilize friction-fit or snap-fit mechanisms designed for permanent stabilization, making them difficult and sometimes impossible to reverse. Joint geometry can instead be designed so that incorrect assemblies are hard to reach in the first place~\cite{park_foolproofjoint_2022}. Standard assembly manuals assume sequential progress and rarely provide disassembly instructions. For BLV individuals, this compounds the penalty of a mistake. Future accessible instruction frameworks must proactively account for this by explicitly explaining the connection mechanics and providing guidance on how to safely disconnect components, transforming irreversible mistakes into manageable setbacks.

\subsubsection{Verification at Closing}

We observed that participants frequently listened to the remaining instructions even after they believed they had completed all steps, to ensure nothing was omitted. This behavior mirrors their physical workspace sweeping to detect leftover parts on the table. These findings suggest that instructions need to describe the final assembled state. Similar to survey knowledge upfront, there is an opportunity to provide verification knowledge at the end. Future work could explore how best to convey this and help users verify the correctness of their assembly similar to the \textit{Assembly Stethoscope}~\cite{liu2025assembly}, but targeted to BLV individuals and using audio-tactile instruction.

\subsection{Beyond Assembly Instructions}
We observed that most participants in our study had little prior assembly experience. Inaccessible instruction is one major reason. However, small components and inaccessible tools are further barriers. For example, five of eight participants struggled with using an Allen key to tighten bolts, and participants also dropped tools and components, and struggled to hold parts in alignment while working. Assembly could be made much more accessible (both for BLV individuals, and for individuals with motor disabilities, or really anyone who assembles furniture) if manufacturers were to rethink joinery and tool usage~\cite{park_foolproofjoint_2022, chang2021daedalus}. This is beyond the scope of this paper, but would be important to consider as furniture designers create new products.

Finally, our work suggests that accessible design can transform the emotional landscape of DIY tasks. To many BLV individuals, assembly is frequently associated with “fear of failure”. We observed participants gaining a boost in confidence when they successfully assembled furniture. This increased sense of agency, and the satisfaction that came with it, points to a value in accessible instruction that goes beyond task completion.

\subsection{Design Recommendations for Assembly Instructions}
Based on our study and discussion, we identify the following opportunities to improve the design of BLV assembly instructions. We do not consider this to be an exhaustive list but rather a concrete set of insights derived from our formative study, qualitative evaluation, and reflection:

\begin{enumerate}
    \item Allow users to navigate efficiently between structural roadmap, procedural instruction, and verification.
    \item Use consistent language across roadmap and procedural instruction.
    \item Anticipate that users will make mistakes during assembly. Instructions should provide a clear "undo" path by explaining how to disconnect joints/parts.
    \item Take a cross-sensory approach when converting visual instruction, taking advantage of the tactile and even auditory properties of materials.
    \item Do not just describe (tactile) features on parts that are functional but also include unused or nonfunctional features that users may encounter by touch.
    \item Avoid relying on complex alignment of tools or parts, and when unavoidable, take advantage of users' proprioceptive sense and body features to support alignment.
    \item On demand, let users query the part list or the part count for the current stage, to verify that they are on track. 
    \item Users' hands are occupied by the assembly, so provide ways to control the instructions by voice or another modality.

\end{enumerate}

\section{Limitations}
\label{sec:limitations}

Assemblies at the scale of a wardrobe or a bed frame raise part count, component weight, and the cost of an error together, and we expect the embodied strategies and the error recovery difficulties we observed to matter more at that scale rather than less; testing this is the direction we would pursue first. Moreover, a researcher was present throughout and intervened when participants could not proceed, and participants worked from instructions that the research team had already verified through the pipeline's human-in-the-loop steps. How people fare unaided, with output the pipeline produced unattended, is therefore untested, and a deployment in participants' own homes would answer both questions at once. Because no comparable accessible instructions exist to serve as a baseline, our evidence that the roadmap helped is observational, resting on the contrast between participants who attended to it and those who skipped it; a controlled comparison against procedural instructions alone would isolate what the roadmap contributes.

\section{Conclusion}

In this work, we explored how to empower BLV individuals to independently assemble flat-packed furniture whose assembly has traditionally relied on inaccessible visual instructions. To study this, we used \name{} as a technology probe that converts existing visual instructions into accessible text-based manuals that cover both high-level structural roadmaps and procedural instructions. Through an in-depth study with eight BLV participants, we observed that successful non-visual assembly requires more than direct textual translation. A structural roadmap provides structural understanding prior to physical assembly. Additionally, our observation reveals under-explored physical strategies, such as using the body to stabilize parts and obtain continuous haptic feedback. We also identified critical points in error recovery, highlighting the need for instructions to address disassembly procedures. Finally, we present eight design recommendations to inform future non-visual instruction frameworks. We envision a future where complex assembly tasks are supported by accessible structural roadmaps and procedural instructions, allowing BLV individuals to assemble furniture independently.

%TC:ignore

% \begin{description}
% \item[\texttt{sidebar}:]  Place formatted text in the margin.
% \item[\texttt{marginfigure}:] Place a figure in the margin.
% \item[\texttt{margintable}:] Place a table in the margin.
% \end{description}

%%
%% The acknowledgments section is defined using the "acks" environment
%% (and NOT an unnumbered section). This ensures the proper
%% identification of the section in the article metadata, and the
%% consistent spelling of the heading.
\begin{acks}
To Robert, for the bagels and explaining CMYK and color spaces.
\end{acks}

%%
%% The next two lines define the bibliography style to be used, and
%% the bibliography file.
\bibliographystyle{ACM-Reference-Format}
\bibliography{thijs-papers, reference}

%%
%% If your work has an appendix, this is the place to put it.
\appendix

\section{Initial Instructions}
\label{appendix:initial_instructions}
1. The side frame
These are the largest wooden parts. They look like the lower letter "h".  Imagine these are the two slices of bread in a sandwich. Everything else will be held between them. They provide the legs that touch the floor and the tall posts that support your back.
Tactile Signature: You are holding a side frame. It contains two legs. They are connected by two rails. The top of the h we later will install a backseat. Run your hands over the frame. You will find most one side has multiple small holes. We will use this side to connect cross pieces. 
2. The Backrest
A narrow rectangular panel made of two vertical bars with three evenly spaced horizontal bars fixed between them. It works like a small ladder-shaped structure that connects the tops of the chair’s side frames and provides support when you lean back.
Tactile Signature: It feels like a short section of a ladder. You’ll find two thicker vertical bars and three thin bars running across between them. At the top end of each vertical beam is one dowel hole for attaching it to the chair.
3.  The Seat Rails
Hold the two wooden rails and feel their length.
The long rail spans between the front legs to create support for a wider seat area, while the short rail connects the back legs.
Each rail has small pilot holes on its wide face for attaching plastic connectors, and dowel holes on the cut ends for joining into the legs.
Together, these rails tie the two h-frames into a rigid square base that the seat panel will later attach to.
4. The Seat
A large, flat board. It is wider at the front and narrower at the back. One side is very smooth. The other side may feel rougher or have pilot holes.

\section{Benchmark Dataset Details}

To support reproducibility and provide a detailed view of the source material used in our evaluation, we have provided the full benchmark dataset at the following anonymized repository:

\begin{center}
    \url{https://anonymous.4open.science/r/assemblyai---human-in-the-loop-assistant-7463/}
\end{center}

\label{apx:benchmark}
\begin{table*}[t]
\AccessibleTableDescription{Detailed inventory of the 34 benchmark manuals, identified as manual\_01 through manual\_34. Columns list furniture type, manufacturer, manual identifier, main-part count, page count, and assembly-step count. Reported main-part counts total 296 and range from 2 to 32 per manual. Manuals contain 1 to 28 pages and 1 to 19 assembly steps. The Autonomier shoe-rack manual has the largest main-part count, 32. The RoyalCraft sideboard manual has the most pages, 28, and assembly steps, 19.}
\centering
\begin{tabular}{llcccc}
\toprule
\textbf{Furniture Type} & \textbf{Manufacturer} & \textbf{Manual ID} & \textbf{Main Parts} & \textbf{Pages} & \textbf{Steps} \\
\midrule
Bookshelf & NicBex & manual\_01 & 9 & 12 & 11 \\
Chair & Kyajah & manual\_02 & 3 & 3 & 6 \\
Office Chair & Zesthouse & manual\_03 & 15 & 2 & 8 \\
Table & BoConcept & manual\_04 & 3 & 2 & 4 \\
Bar Stool & BoConcept & manual\_05 & 2 & 2 & 2 \\
Chair & BoConcept & manual\_06 & 2 & 1 & 1 \\
Dining Chair & BoConcept & manual\_07 & 2 & 2 & 5 \\
Bench & BoConcept & manual\_08 & 3 & 4 & 8 \\
Chair & BoConcept & manual\_09 & 5 & 6 & 9 \\
Bench & Furniture of America & manual\_10 & 5 & 5 & 2 \\
Shelf Unit & Furniture of America & manual\_11 & 7 & 5 & 4 \\
Ottoman & Furniture of America & manual\_12 & 7 & 7 & 4 \\
Cart & Furniture of America & manual\_13 & 5 & 7 & 4 \\
Bookcase & nathanjames & manual\_14 & 7 & 3 & 9 \\
Table & Modway & manual\_15 & 4 & 9 & 5 \\
Cabinet & DSHADE & manual\_16 & 29 & 11 & 15 \\
Sideboard & RoyalCraft & manual\_17 & 14 & 28 & 19 \\
Cart & O\&K & manual\_18 & 17 & 2 & 6 \\
Chair & CLATINA & manual\_19 & 4 & 2 & 4 \\
Floor Lamp & CIOC & manual\_20 & 3 & 4 & 2 \\
Floor Lamp & CIOC & manual\_21 & 12 & 7 & 4 \\
Ottoman & DREAMS IDEA & manual\_22 & 7 & 12 & 4 \\
Desk Chair & Sweetcrispy & manual\_23 & 9 & 2 & 4 \\
Side Table & GKAN & manual\_24 & 8 & 2 & 5 \\
Side Table & Highland Dunes & manual\_25 & 5 & 4 & 3 \\
Storage Bench & KaBuea & manual\_26 & 10 & 14 & 7 \\
Bookcase & VASAGLE & manual\_27 & 12 & 12 & 5 \\
Table & DELLMAO & manual\_28 & 8 & 8 & 3 \\
Hanging Chair & Inspire & manual\_29 & 6 & 2 & 8 \\
Nightstand & Furinno & manual\_30 & 11 & 4 & 3 \\
Office Chair & TECHNI MOBIL & manual\_31 & 9 & 7 & 3 \\
Shoe Rack & Autonomier & manual\_32 & 32 & 1 & 5 \\
Armchair & Yaheetech Store & manual\_33 & 8 & 12 & 3 \\
Office Chair & BOWTHY & manual\_34 & 13 & 2 & 7 \\
\bottomrule
\end{tabular}
\caption{Summary of tested furniture assembly manuals detailing furniture type, manufacturer, assigned identifier, and key layout metrics.}
\label{tab:furniture_manuals}
\end{table*}

\section{Prompts}\label{appendix:prompts}

\subsection{Step 1: Structural Part Identification}
\begin{taglstlisting}
`You are an expert at analyzing furniture assembly manuals. Your task is to identify all the MAJOR structural parts of the furniture.

I have provided multiple images, each labeled with an "Image Index" (e.g., Image Index 0, Image Index 1).

TASK:
1. Find all occurrences of structural parts (Seat, Backrest, Legs, Armrests, Wheels, etc.).
2. For each part, identify which image contains the clearest "Parts Catalog" view (usually a clean diagram of the part alone with its label) vs an "Assembly View" (the part being put together).
3. Select ONE Primary detection for each part---this should be the clearest, most isolated view of the part (usually from the components list on the first page).

CRITICAL CONSTRAINTS:
- Identify ONLY major structural parts.
- IGNORE small hardware (bolts, screws, washers) and tools (wrenches).
- If a part appears in multiple images, list all occurrences but mark the BEST one as "isPrimary": true.
- BOUNDING BOX ENFORCEMENT: The bounding box (box_2d) MUST tightly surround the graphical drawing/illustration of the part itself. Do NOT put the bounding box over or around the text label. The box should encompass the picture of the furniture part.
- SYMMETRY/SEGMENTATION: If a single label (e.g., "8 (L&R)", "1 x2") points to multiple physically separate items (e.g., two distinct leg frames, or a left armrest and a right armrest), you MUST draw a separate bounding box and create a separate JSON object for EACH item. Name them distinctly (e.g., "Left Armrest" and "Right Armrest").
- ALTERNATIVES: If the manual shows alternative versions of a part for different models, only extract one primary baseline version.

Return the result ONLY as a JSON array of objects using normalized format for coordinates (0-1000 scale where 0 is top/left and 1000 is bottom/right):
[
  {
    "name": "Part Name",
    "label": "A",
    "pageIndex": 0,
    "box_2d": [ymin, xmin, ymax, xmax],
    "isPrimary": true
  }
]`
\end{taglstlisting}

\subsection{Step 2: Material Selection and Mapping}
\begin{taglstlisting}
`You are a Blind, low vision assistant responsible for guiding assembling flat packed furnitures. Your inputs include {A}: an image of the scene consisting of furniture parts labeled with numbers/letters, {B}: a JSON file that describes the image's objects and labels, and {C}: a set of furniture setup manual pages.

First, you are ONLY responsible for identifying the relevant materials that will be required to assemble the furniture in the image. Output a table of selected materials, with their labeled numbers, and a brief explanation of why they are selected and how are they related to items on the set up manual. The table format should be json, and it should be really similar to {B}, but with an additional explanation section for each selected material and their labeled number. Hint: Usually in 99.999% of cases, the number of selected materials equals to the numbered of labeled furniture parts.

TASK:
Analyze how the identified parts connect and function based on the provided manual and images.
For each material, locate its position(s) on the most fully assembled view where the part is still visible (this could be the final assembly on the last page, or a partially assembled frame in an earlier step). Provide the \`targetArea\`: \`{"pageIndex": number, "points_2d": [[y1, x1], [y2, x2]]}\` mapping the exact pixel centers of where these parts belong (using a 0-1000 normalized scale). If there are multiple instances of a part (e.g., 4 legs), provide a coordinate array covering each visible instance. Omit if uncertain.
Additionally, you must evaluate a \`confidenceScore\` ("High", "Medium", or "Low") for your targetArea mapping and an accompanying \`confidenceJustification\` for why you graded it that way. Output "Low" if you cannot confidently trace the target connections or if the part is blurry/ambiguous.

CRITICAL VERIFICATION: Before finalizing your output, double check:
1. Ensure multiple points are not redundantly pointing to the exact same physical instance of a part.
2. Ensure one physical part on the image is not accidentally being referenced by two different part labels.
3. Verify that the coordinates actually land accurately on the intended item in the visual diagram.
4. If a part is completely internal or hidden in the final assembled view, use an earlier manual page (\`pageIndex\`) where it is visibly being installed rather than guessing a location on the final page.

Return the result ONLY as a JSON array of objects:
[
  {
    "name": "Part Name",
    "label": ["A"],
    "explanation": "Brief explanation of why this was selected and how it relates to items on the setup manual.",
    "confidenceScore": "High",
    "confidenceJustification": "The part is clearly visible on the final assembly diagram and its shape matches perfectly.",
    "targetArea": {
      "pageIndex": 0,
      "points_2d": [[500, 500], [520, 520]]
    }
  }
]`
\end{taglstlisting}

\subsection{Step 2: Re-verification}
\begin{taglstlisting}
`You are an AI tasked with writing short, clear function descriptions of furniture parts based on their visual connections.

I have provided multiple images of the furniture setup manual pages.
I have also provided a JSON file containing the user's VERIFIED connection locations for each part. 
The "targetArea" contains the exact manual page (pageIndex) and pixel coordinates (points_2d) where the user has determined this part ultimately attaches.

TASK:
Based on the provided location where the part is connected in the manual images, write a brief explanation describing how this part functions and attaches to the rest of the furniture structure. Keep it under 2 sentences.

Ensure your JSON output maintains the exact same structure as the input, just updating the "explanation" strings. DO NOT change the names, labels, or targetAreas.

Return the result ONLY as a JSON array of objects.
`
\end{taglstlisting}

\subsection{Step 3: Structural Roadmap Generation}
\begin{taglstlisting}
`You are writing an audio-first, non-visual "mental model roadmap" for blind/low-vision furniture assembly. This description should help people to understand how this furniture will be built by connecting each part. By following your description, blind people can use clay make the 3d model right at least 90% right. For simplify consideration, you can choose not overly describe some complex part structure in details but you should provide how do that overall look like or its function.


For your inputs:
You will receive assembly manual images. This description will include graphic assembly instructions, and on the left of some sections/pages, there might be a big black bolded number, which indicates the current assembly step number.
You will also receive a JSON formatted object named parts_connections_data that describes how parts connect: {parts_connections_data}. Note that parts_catalog.png is an image containing part images and labels, but you should rely on the provided JSON data for part information, not directly on the image.

Goal: Help the user form a correct structural mental model BEFORE assembly

Hard constraints:
Use simple, concrete language.
Your output will be played to blind people, bracket, bold words, colon and dash have no meaning to them.
Avoid visual metaphors (no "looks like a letter," no "imagine a picture").
Avoid exact measurements in cm/inches unless absolutely necessary; prefer relative anchors (seat height, mid-height, near the floor).
Use consistent part names. If two parts are similar, distinguish by position (front seat rail vs back rail) not by vague adjectives (shorter/longer).
Keep the mental model concise(<300 words).
Do not refer to parts using wayfair labels like 'part A,' 'part B,' 'part C,' or 'part D,' as these are meaningless to blind users. Always use the consistent part names provided.
Do not explain the assembly of small fasteners like individual bolts, screws, or nuts. Focus on the connection of major furniture components.

########## assistant example start ##########
You are building a bookcase without doors or drawers.
All main wooden parts arrive as separate flat panels with holes already drilled, and the back arrives as a separate thin sheet, so no large section is pre-assembled.
There are seven main structural units: left side panel, right side panel, bottom panel, top panel, fixed shelf panel, thin back panel, and four adjustable shelves with shelf pins.
The left and right side panels are the two tall uprights, and every other panel spans between them.
The bottom panel locks the uprights together near the floor and forms the lowest flat support surface.
The fixed shelf locks the uprights together again around mid height and is the key internal brace that reduces twisting.
The top panel locks the uprights together at the top edge and completes a tall open box with an open front and an open back.
The thin back panel closes the back opening and stops falling from the back.
Each adjustable shelf is a removable flat surface that rests on four shelf pins,each one near corners, so shelves can move up or down without changing the frame.
########## assistant example end ##########

Now it is your turn for generating a mental model description based on the provided inputs.

`
\end{taglstlisting}

\subsection{Step 4: Step-by-Step Instructions}
\begin{taglstlisting}
`You are writing step-by-step assembly instructions for blind/low-vision users. These instructions will be read aloud immediately after a structural overview of the furniture.

For your inputs:
You will receive assembly manual images. These include graphic assembly instructions, and on the left of some sections/pages, there might be a big black bolded number, which indicates the current assembly step number.
You will receive a JSON formatted object named parts_connections_data that describes the verified part connections: {parts_connections_data}.
You will also receive the mental model description that was already read to the user: {mental_model_text}. You MUST use the exact same part names established in the mental model so the user does not get confused.

TASK:
Write the sequential assembly instructions by reading the manual's numbered assembly steps from the images. For each step:
State the step number.
Name the parts being connected in that step, using the exact same part names from the mental model.
Describe what action to perform and how the parts join together.
Mention orientation cues where helpful, such as which side faces up, which edge faces front, or which hole to align.
Describe individual bolt or screw insertion in detail. Ensure you explain how small parts such as bolts are installed. Instead of just saying "secure the seat to the frame", explicitly say "insert bolt through the rear of the seat under and connect it with the backrest."
If the manual shows the same step from multiple angles, merge them into one clear instruction.

Hard constraints:
Use simple, concrete language.
Your output will be played to blind people. Brackets, bold words, colons, dashes, and bullet points have no meaning to them. Write in plain flowing sentences.
Avoid visual metaphors such as "looks like a letter" or "imagine a picture."
Avoid exact measurements in centimeters or inches unless absolutely necessary. Prefer relative anchors such as seat height, mid height, or near the floor.
Use the exact same consistent part names from the mental model throughout. Do not introduce new names for parts already described.
Do not refer to parts using wayfair labels like part A, part B, part C, or part D.
Do not use any markdown formatting, headers, or special characters. Write everything as plain readable sentences and paragraphs.
Begin your output directly with "Now let us begin the assembly." followed by the first step.

########## assistant example start ##########
Now let us begin the assembly.

Step one. Lay the left side panel flat on the ground with the pre-drilled holes facing up. Position the bottom panel so its left edge aligns with the bottom holes of the left side panel. Secure the bottom panel to the left side panel.

Step two. Attach the fixed shelf panel to the left side panel at the mid height holes. This shelf will act as the main brace that prevents the bookcase from twisting.

Step three. Place the right side panel onto the exposed ends of the bottom panel and the fixed shelf panel. Secure both connections. You now have a tall open frame standing on its own.

Step four. Lower the top panel onto the top edges of both side panels and secure it. The basic box structure is now complete.

Step five. Lay the thin back panel against the open back of the frame. Fasten it along all four edges. This panel squares the frame and prevents racking.

Step six. Insert four shelf pins into the rows of holes at your desired shelf heights. Rest each adjustable shelf on its four pins.
########## assistant example end ##########

Now it is your turn. Generate the step-by-step assembly instructions based on the provided inputs.

`
\end{taglstlisting}
\section{Instructions Provided to Participants}\label{appendix:instructions}

The following are the exact text instructions generated by the
human-in-the-loop pipeline. Participants accessed these instructions using screen readers during the user study.

\subsection{Shoe Rack Assembly Instructions}

You are building a four tier shoe rack. The structure is completely modular and built using tubes and plastic joints without any screws. 

The overall structure consists of four identical flat shelf layers stacked vertically on top of each other. 

First let us understand a single shelf layer. Each layer is made of one flat fabric shelf, two long tubes, and two plastic connectors. The fabric shelf has open sleeves along its front and back edges. The horizontal long tubes slide entirely through these sleeves. This gives the soft fabric shelf a rigid front edge and a rigid back edge. 

To complete one shelf layer, the left and right plastic connectors plug into the open ends of the long tubes. These plastic connectors form the hard left and right side borders of the shelf layer. You will build four of these exact same flat layers.

Once your shelf layers are built, you will stack them to create the full height of the rack. This is where the short tubes are used. The short tubes act as vertical support pillars. They plug directly into holes on the top surface of the plastic connectors of the bottom shelf layer, standing straight up near the four corners. The next shelf layer is then placed directly on top, plugging into these upright short tubes. 

You will repeat this stacking process using the remaining short tubes to connect all four layers. The bottom layer rests directly on the floor, and the top layer forms the highest open shelf. The final result is a freestanding shelving unit.

Now let us begin the assembly.

Step one. Take one fabric shelf. Next, plug the left ends of both long tubes into the side holes of one plastic connector. Then, plug the right ends of the long tubes into the side holes of the second plastic connector. Ensure the fabric is stretched flat. Repeat this entire process three more times so that you have built a total of four identical flat shelf layers.
 
Step two. Place one of the completed shelf layers flat on the floor to serve as the bottom tier. Take four short tubes. Insert one short tube into each of the four vertical holes located on the top surface of the left and right plastic connectors. Press the tubes down firmly so they stand straight up like pillars.

Step three. Take a second completed shelf layer. Align the holes on the underside of its plastic connectors with the four upright short tubes on your bottom layer. Push the second shelf layer down firmly onto the short tubes. Next, insert four more short tubes into the top holes of this newly added second layer.

Step four. Take a third completed shelf layer and place it onto the upright short tubes from the previous step. Press it down firmly to secure it. Then, insert your final four short tubes into the top vertical holes of this third layer.

Step five. Take your fourth and final completed shelf layer. Align the holes underneath its plastic connectors over the final set of upright short tubes. Press down firmly on all four corners to secure the top tier. Your four tier shoe rack is now completely assembled.

\subsection{Desk Chair Assembly Instructions}

You are building a simple rolling office chair without armrests. There are five main structural components. These are the base, five casters, a gas lift, a seat, and a back. 

The chair is built in two main sections that join at the very end.

The bottom section is the rolling support. The base sits near the floor and has five legs spreading outward from a central hub. One caster plugs into the bottom tip of each leg to act as a wheel. The gas lift is a vertical cylinder that drops into the top center of the base. This cylinder holds up the chair and adjusts the height.

The top section is the sitting area. The seat is the main flat horizontal surface. The back attaches securely under the rear edge of the seat and curves straight up. 

To finish the structure, the connected seat and back drop directly onto the top of the vertical gas lift. This firmly links the top sitting area down to the bottom rolling base.

Now let us begin the assembly.

Step one. Place the base upside down on the floor so the bottom of the legs face upward. Press one caster firmly into the hole located at the bottom tip of each of the five legs. You will use all five casters.

Step two. Turn the base right side up so the newly installed casters rest flat on the floor. Take the gas lift and drop its wider bottom end directly into the open center hole of the base. It will slide in and rest firmly in place. This completes the rolling support section.

Step three. Lay the seat upside down on a flat surface so the bottom faces up. Bring the lower extension of the back against the rear edge of the seat so their connection holes line up. Insert the two bolts horizontally through the holes in the lower extension of the back and into the threaded holes of the seat. Use the included wrench to turn each bolt clockwise until they are completely tight. You now have the sitting area assembled.

Step four. Pick up the assembled sitting area and turn it right side up. Position it directly over your rolling support section. Align the center hole underneath the seat with the top of the gas lift. Lower the seat straight down onto the gas lift and press down firmly on the top of the seat to securely lock the top and bottom halves together. The chair assembly is now complete.

\section{Semi-Structured interview questions}\label{appendix:questions}
The following questions constitute our interview guide. While these questions were asked of all participants, we frequently asked follow-up questions to probe deeper into specific perspectives raised during the conversation.
\subsection{User Experience and Support}
\begin{itemize}
    \item What aspects of the system worked well during your assembly?
    \item Were there moments when you felt confused or insufficiently supported?
    \item Can you describe a specific instance where you felt stuck? What information was missing for you to succeed?
    \item Is there more the system could have done to support you throughout the process?
\end{itemize}

\subsection{Information Granularity and Control}
\begin{itemize}
    \item Did you feel the spoken instructions were appropriately detailed?
    \item Did you feel in control of the assembly process, or did you feel ``led'' by the system?
    \item Did you feel you had a clear mental model of your progress? Can you provide examples of where this overview broke down?
\end{itemize}

\subsection{Accessibility and Future Improvements}
\begin{itemize}
    \item How well does this workflow align with your existing assistive technology (e.g., screen readers)?
    \item Do you feel confident in the integrity of your final assembly?
    \item If we could improve one specific feature of this workflow, what should we prioritize?
\end{itemize}

%TC:endignore
\end{document}